\documentclass{aa}
\usepackage{graphicx}
\usepackage{txfonts}
\usepackage{upgreek}
\usepackage{natbib}
\usepackage{xcolor}
\usepackage{amsmath}
\usepackage[colorlinks=true,linkcolor=blue,citecolor=blue,urlcolor=blue]{hyperref}

\begin{document} 

\title{JOYS: Linking the molecular ice and gas-phase composition towards the high-mass hot core IRAS\,18089$-$1732}
\author{C. Gieser
	\inst{1}
	\and
	W.~R.~M. Rocha\inst{2,3}
	\and
	Y. Chen\inst{3}
	\and
	K. Slavicinska\inst{2}
	\and
	E.~F. van Dishoeck\inst{3,4}
	\and
	P. Nazari\inst{5}
	\and
	N.~G.~C. Brunken\inst{3}
	\and
	L. Francis\inst{3}
	\and
	H. Beuther\inst{1}
	\and
	S. Reyes-Reyes\inst{1}
	\and
	A. Caratti o Garatti\inst{6}
	\and
	P.~D. Klaassen\inst{7}
	\and
	J.~M. Vorster\inst{8}
	\and
	M. G. Navarro\inst{9}
	}

	\institute{
	Max Planck Institute for Astronomy, K\"onigstuhl 17, 69117 Heidelberg, Germany\\
	\email{gieser@mpia.de}
	\and
	Laboratory for Astrophysics, Leiden Observatory, Leiden University, PO Box 9513, NL 2300 RA Leiden, The Netherlands
	\and
	Leiden Observatory, Leiden University, PO Box 9513, NL 2300 RA Leiden, The Netherlands
	\and
	Max Planck Institut für Extraterrestrische Physik (MPE), Giessenbachstrasse 1, 85748 Garching, Germany
	\and
	European Southern Observatory, Karl-Schwarzschild-Strasse 2, 85748 Garching, Germany
	\and
	INAF – Osservatorio Astronomico di Capodimonte, Salita Moiariello 16, 80131 Napoli, Italy
	\and
	UK Astronomy Technology Centre, Royal Observatory, Edinburgh, Blackford Hill, Edinburgh, EH9 3HJ, United Kingdom
	\and
	Department of Physics, P.O. box 64, FI- 00014, University of Helsinki, Finland
	\and
	INAF – Osservatorio Astronomico di Roma, Via di Frascati 33, 00078 Monte Porzio Catone, Italy
	}

	\date{Received x; accepted x}

	\abstract
	{The formation and destruction of molecules in the interstellar medium is a complex interplay between gas-phase reactions as well as processes on grain surfaces and within icy mantles. For many decades, the gas-phase composition of the cold material towards star-forming regions could be well characterized using (sub)mm facilities. Prior to the launch of the James Webb Space Telescope (JWST), ice species other than the main constituents (H$_{2}$O, CO, CO$_{2}$, NH$_{3}$, CH$_{4}$, CH$_{3}$OH) were challenging to detect due to insufficient sensitivity as well as angular and/or spectral resolution.}
	{We determine molecular ice and gas-phase column densities towards the young and embedded high-mass hot core IRAS\,18089$-$1732 within a region of 5\,000\,au.}
	{We use spectroscopic data from 5-28\,$\upmu$m obtained with JWST to derive ice column densities of H$_{2}$O, SO$_{2}$, OCN$^{-}$, CH$_{4}$, HCOO$^{-}$, HCOOH, CH$_{3}$CHO, CH$_{3}$COOH, C$_{2}$H$_{5}$OH, CH$_{3}$OCH$_{3}$, and CH$_{3}$COCH$_{3}$. Gas-phase column densities of a total of 38 molecules, including, O-, N-, S-, and Si-bearing species as well as less abundant isotopologues, are inferred from sensitive molecular line observations taken with the Atacama Large Millimeter/submillimeter Array (ALMA) at 3\,mm wavelengths.}
	{We find comparable abundances (relative to C$_{2}$H$_5$OH or CH$_{3}$OH) in both phases for C$_{2}$H$_5$OH, CH$_{3}$OH, and CH$_{3}$OCH$_{3}$. The abundances of SO$_{2}$ and CH$_{3}$COCH$_{3}$ are higher in the gas-phase suggesting additional gas-phase formation routes. The abundance of CH$_{3}$CHO is one order of magnitude higher in the ices compared to the gas-phase. The ice abundances (relative to H$_{2}$O ice) towards the IRAS\,18089 hot core are similar to previously studied Galactic low- and high-mass protostars. There are hints of a decreasing abundance with Galactocentric distance for OCN$^{-}$, CH$_{3}$OH, and CH$_{3}$CHO ice.}
	{Not all species show comparable abundances in the ice and gas-phases. However, we find similar trends when species show elevated ice or gas-phase abundances in the high-mass hot core IRAS\,18089 compared to low-mass hot cores. To better understand the reaction pathways of molecular species, statistical surveys analyzing both the ice and gas-phase chemical composition of high- and low-mass protostars at different Galactocentric radii are essential.}
	
	\keywords{Stars: formation -- Stars: massive -- Stars: protostars -- ISM: molecules -- ISM: individual objects: IRAS 18089-1732 }
	\maketitle
	\nolinenumbers

\section{Introduction}\label{sec:intro}

\begin{figure*}[!htb]
\centering
\includegraphics[width=0.99\textwidth]{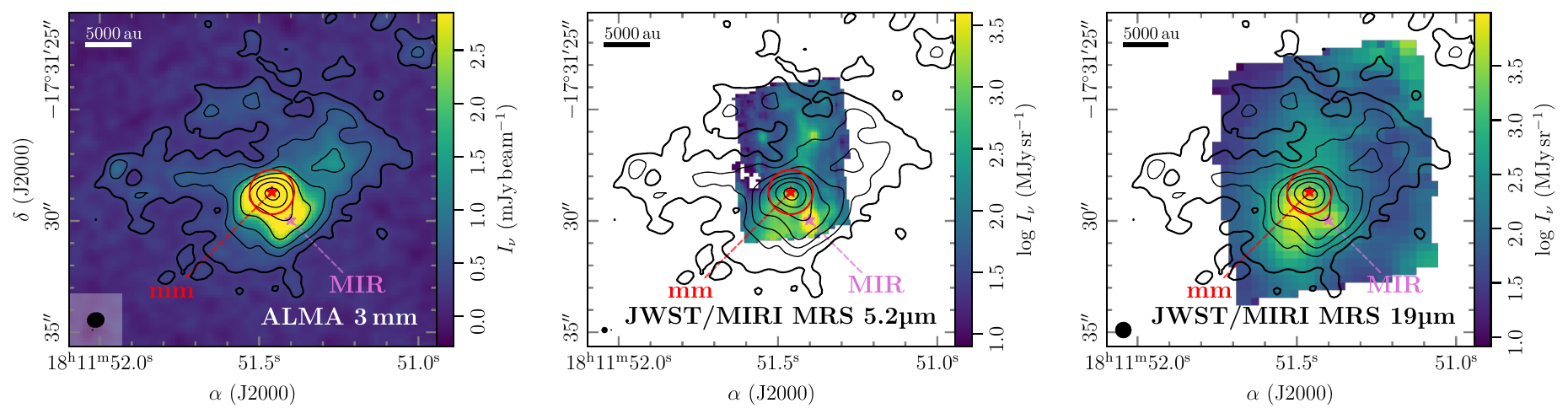}
\caption{IRAS18089 continuum images (left: ALMA 3\,mm, center: JWST/MIRI-MRS 5$\upmu$m, right: JWST/MIRI-MRS 19$\upmu$m). In all panels the black contours are the ALMA 3\,mm continuum with steps from 5, 10, 20, 40, 80, 160, 320$\times\sigma_\mathrm{cont}$. The mm and MIR continuum peak positions are labeled and highlighted in red and pink. The red circle shows the aperture (1$''$ radius) used for spectra extraction towards the mm source. A scale bar in the top left panel marks a spatial scale of 5\,000\,au. The ellipse in the bottom left corner highlights the angular resolution of each data set.}
\label{fig:continuum}
\end{figure*}

	During the earliest stages of high-mass star formation, protostars are still deeply embedded in, and accreting from the surrounding cloud material \citep{Motte2018}. Within star-forming clouds, massive star formation is commonly observed towards dense clumps, with typical sizes of $\sim$0.1-1\,pc. These clumps are often located in hub-systems, which are converging points of large-scale filaments \citep{Kumar2020}. Within a single clump, smaller cores ($<$0.1\,pc) are embedded where each core is expected to form a single or small multiple system of stars \citep{Beuther2018,Sanhueza2019,Gieser2023a,Louvet2024}. 
	
	The classification of evolutionary stages during massive star formation is motivated by multi-wavelength observations from infrared (IR) to radio wavelengths \citep{Beuther2007}. Infrared dark clouds mark the youngest phase where the cloud is dominated by very cold gas and dust \citep{Rathborne2006}. As cores collapse and protostars reach their main accretion phase, the high-mass protostellar object (HMPO) stage is reached \citep{vanderTak2000,Sridharan2002,Beuther2002,Shirley2003}. High luminosities and energetic outflows are observed as well as a shift of the peak spectral energy distribution (SED) towards the mid-infrared (MIR) \citep{Beuther2002b} indicating high accretion rates \citep[$\sim$10$^{-4}$\,$M_\odot$\,yr$^{-1}$,][]{deWit2010,Beltran2016,DeBuizer2017,Lu2018}. Spectra with many molecular lines are observed at (sub)mm wavelengths towards high-mass star-forming regions (HMSFRs) when gas temperatures above 100\,K are reached, due to the sublimation of molecular ices from dust grains into the gas-phase \citep{Garrod2006,Herbst2009}. This stage is commonly referred to as the hot (molecular) core phase \citep{Cesaroni1997}. Eventually hyper-/ultra-compact H{\sc ii} regions form due to the high UV luminosity of the central (proto)stars which ionize and eventually disrupt the surrounding envelope material \citep{Churchwell2002,Dale2012,Klaassen2018}.

	Over the last decades, the gas-phase composition towards HMSFRs obtained by sensitive radio telescope and interferometers have revealed a great complexity as a function of evolutionary stage as well as Galactic radius \citep{Gerner2014,Woods2015,Coletta2020,Gieser2021,Sabatini2021,vanGelder2022,Chen2023,Bouscasse2024,Gigli2025,Nazari2025}. Contrary to that, only a few ice species were detected towards HMSFRs, mostly using the Short Wavelength Spectrometer on the Infrared Space Observatory \citep{Ehrenfreund1998,Dartois1999,Schutte1999,Gibb2000,Gibb2004,Boogert2015,Boogert2022}. Previous ice studies towards HMSFRs were limited by poor angular resolution, sensitivity and/or spectral resolution. This prevented so far a detailed investigation of the ice composition towards individual protostellar cores. Based on the gas-phase inventory of HMSFRs, chemical models predict that grain-surface chemistry on dust grains is important for the formation of complex organic molecules \citep[COMs,][]{Garrod2006,Herbst2009}.
	
	A direct comparison between gas-phase and ice abundances have been carried out only towards a few low-mass protostars \citep{Perotti2021,Chen2024}. It is therefore crucial to study also high-mass protostars which evolve on shorter timescales, are embedded in clustered regions, and heat up their surrounding envelope material that can all heavily influence the chemical properties of the molecular gas. Studying these environments is important since most stars form in dynamically active HMSFRs, as opposed to the nearby isolated low-mass star-forming regions.
	
	With the launch of the James Webb Space Telescope (JWST), more sensitive and higher spectral and spatial resolution observations in the mid-infrared (MIR), that allows a detailed characterization of interstellar ices, are now possible \citep{Yang2022, McClure2023,Rocha2024, Chen2024, Nazari2024, Slavicinska2024, Brunken2024, Slavicinska2025, Nakibov2025, Smith2025, Rocha2025, Rayalacheruvu2025, Sewilo2025, Tyagi2025, Gross2026}. Most importantly, the detection of COMs other than CH$_{3}$OH can now be achieved \citep{Rocha2024}. While most of the evolved high-mass protostars in the Milky Way are too bright for JWST instruments, deeply embedded HMPOs can still be targeted due to their high IR extinction.
	
	The JWST Observations of Young protoStars (JOYS) program (PI: E. F. van Dishoeck) targets a sample of low- and high-mass protostars using the Mid Infrared Instrument and Medium resolution Spectrograph (MIRI-MRS), and providing spectroscopic information from 5 to 28\,$\upmu$m. An overview of the observations and science results is presented in \citet{vanDishoeck2025}. These MIR observations of star-forming regions with JWST allow a detailed analysis of atomic, ionic and molecular gas-phase lines as well as molecular ices \citep{vanDishoeck2025}.
	
	IRAS\,18089$-$1732 (also commonly referred to as G12.89+0.49, we use IRAS\,18089 hereafter) is a HMSFR in the HMPO stage at a distance of $2.5\pm0.3$\,kpc ($\varv_\mathrm{LSR}=33.8$\,km\,s$^{-1}$) \citep{Urquhart2018}. The clump mass and bolometric luminosity are estimated to be $10^{3.1}$\,$M_\odot$ and $10^{4.3}$\,$L_\odot$, respectively \citep[][]{Urquhart2018}. For a multi-wavelength overview from large to small scales in the region, we refer to Appendix B in \citet{Gieser2023a}. 
	
	Figure \ref{fig:continuum} presents an overview of the hot core region of IRAS\,18089. The region is associated with CH$_{3}$OH maser emission \citep{Beuther2004}, which shows variability with a stable period suggesting the presence of a binary \citep{Goedhart2009}. Both the dust continuum and magnetic field morphology show spiral-like structures towards the central core \citep{Sanhueza2021}. High resolution ($<0.1''$) data with the Atacama Large Millimeter/submillimeter Array (ALMA) show disk-like kinematics that can not be explained by a simple Keplerian rotation model \citep{Ginsburg2023}. The disk-like rotation is roughly perpendicular to the north-south directed molecular outflow oriented near the plane of sky \citep[][]{Beuther2004}. The 3\,mm ALMA continuum reveals several cores surrounding the central hot core and a Spitzer 4.5\,$\upmu$m point source is detected in the vicinity of the central source \citep{Gieser2023a}. 
	
	In the central region of the main core, the temperature exceeds 100\,K within a radius of 10\,000\,au and a steep temperature profile ($T\sim r^{-q}$) with $q$=0.8$\pm$0.1 is inferred from CH$_{3}$CN line emission \citep{Gieser2023a}. At temperatures above 100\,K most COMs that were frozen on the icy grains sublimate into the gas-phase \citep{Garrod2022}. Hence IRAS\,18089, targeted both by sensitive JWST \citep{vanDishoeck2025} and ALMA \citep{Gieser2023a} observations, is the ideal hot core source to study the chemical inventory in the ices as well as in the gas-phase to achieve a better understanding of COM formation processes in HMSFRs. The ice composition towards the MIR source of IRAS\,18089 (Fig. \ref{fig:continuum}) was analyzed in \citet{vanDishoeck2025}, whereas only at the hot core position (IRAS\,18089 mm) both the ice and gas-phase can be directly compared using data sets at similar spatial resolutions.

\section{Observations}\label{sec:observations}

	In the following we describe the reduction of spectroscopic data of IRAS\,18089 in the MIR (5-28$\upmu$m) obtained with JWST (Sect. \ref{sec:JWSTobs}) as well as at 3\,mm wavelengths (86.3-110.3\,GHz) using ALMA (Sect. \ref{sec:ALMAobs}).
	
\begin{figure*}[!htb]
\sidecaption
\centering
\includegraphics[width=0.7\textwidth]{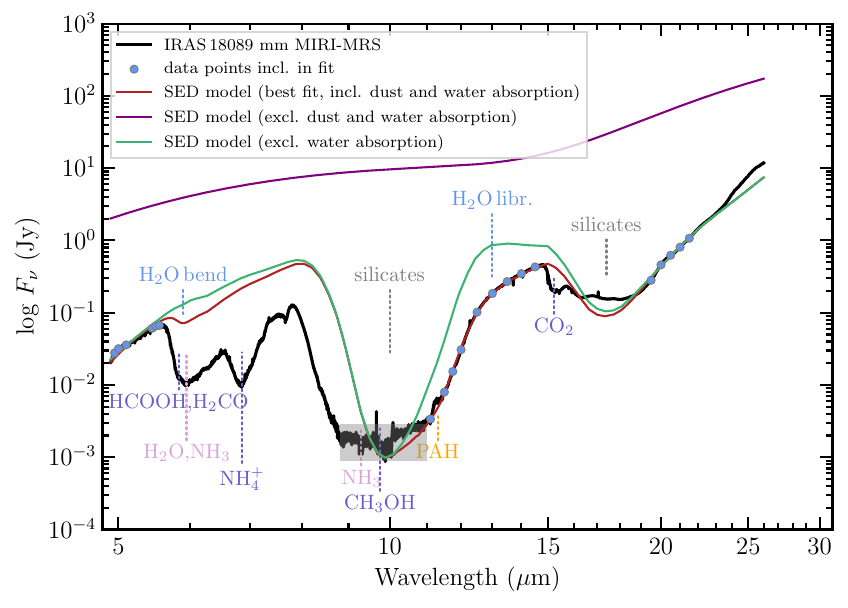}
\caption{JWST/MIRI-MRS spectrum of IRAS\,18089 mm. The observed MIR spectrum of the mm source is shown in black. The red line is the best-fit SED model taking into account emission by two black bodies ($T_1=410$\,K and $T_2=83$\,K) and absorption by dust and water ice. The data points that were included in the fit are highlighted by the blue dots. The green line shows the contribution by dust absorption and the purple line highlights the two black body emission components. Silicate features as well as main ice constituents \citep{Whittet1996, Gibb2000, Yang2022,McClure2023} are labeled. The wavelength range with an increased noise (8.8-11.0\,$\upmu$m) is grey-shaded.}
\label{fig:spectrumIR}
\end{figure*} 

\subsection{JWST/MIRI-MRS}\label{sec:JWSTobs}
	
	The observations with JWST \citep{Gardner2023,Rigby2023} are part of the JOYS program (program ID: 1290, PI: E. F. van Dishoeck). IRAS\,18089 was observed on September 18th 2023 using the MRS mode of the MIRI instrument \citep{Rieke2015,Wright2023} in two consecutive mosaic pointings, as well as a nearby dedicated background pointing ($\alpha_\mathrm{J2000}$=18:11:44.1236 and $\delta_\mathrm{J2000}$=$-$17:31:3.10). The observations included all three grating settings (short, medium, long) and all channels (ch1, ch2, ch3, and ch4) providing in total 12 sub-bands covering a wavelength range from 4.9\,$\upmu$m to 28\,$\upmu$m. The on-source time was 200\,s per grating setting and per pointing. We used the 2-point dither mode that is appropriate for extended sources and the default readout mode FASTR1. Due to spatial and spectral undersampling, small amplitude modulations can be present in extracted MIRI-MRS spectra using a 2-point instead of a 4-point dither \citep[Fig. 9][]{Law2023}. However, since our aperture for spectral extraction is large (Sect. \ref{sec:spectraextraction}), our data quality is not affected by this effect. The dedicated background observation was taken in the same setup with the same integration time. Strong emission lines were masked before background subtraction. A detailed description of the JOYS data reduction pipeline is presented in \citet{vanGelder2024b}. The IRAS\,18089 MIRI-MRS data were calibrated using the same pipeline version \citep[1.13.4,][]{Bushouse2024} and reference context (jwst\_1188.pmap).
	
	The final IRAS\,18089 data products consist of one data cube per sub-band including the combined 2-pointing mosaic. The angular resolution $\theta$ and spectral resolving power ($R=\lambda/\Delta \lambda$) range from 0.2$''$ to 0.8$''$ and 4000 to 1500 from 5 to 28\,$\upmu$m, respectively \citep{Argyriou2023,Jones2023,Law2023}. The 5.2\,$\upmu$m and 19\,$\upmu$m continuum maps (Fig. \ref{fig:continuum}) were estimated between $5.2-5.3$\,$\upmu$m (ch1 short) and $19.0-19.1$\,$\upmu$m (ch4 short) covering wavelength ranges that are absent of strong line emission or absorption.

\subsection{ALMA Band 3}\label{sec:ALMAobs}

	The ALMA observations of IRAS\,18089 were carried out in Cycle 6 at 3\,mm wavelengths (project code: 2018.1.00424.S, PI: C. Gieser). A detailed description of the data reduction and imaging procedure is presented in Sect. 3 and Appendix A in \citet{Gieser2023a}. In this work, we use the 3\,mm continuum as well as the spectral line data of in total 23 spectral windows (spws) covered by two receiver tunings covering different spectral ranges (SPRs, we refer to them as SPR1 and SPR2 in the following).
	
	The ALMA 3\,mm continuum data is presented in \citet{Gieser2023a} and was created by combining continuum channel ranges in the entire 3\,mm data set. The noise level of the 3\,mm continuum is $\sigma_\mathrm{cont}$=0.057\,mJy\,beam$^{-1}$ and the synthesized beam ($\theta_\mathrm{maj}\times\theta_\mathrm{min}$) is $0.79''\times0.70''$ with a position angle (PA) of $101^\circ$. The observational setup of the spectral windows is summarized in Table \ref{tab:alma_obs}. Depending on the spectral setup and frequency, the angular resolution of the spws vary from $\approx0.6''$ to $\approx1.9''$. Three broadband spws have a channel width of 0.977\,MHz ($\approx$3.4\,km\,s$^{-1}$), while the remaining narrowband spws have a channel width of 0.244\,MHz ($\approx$0.9\,km\,s$^{-1}$). The line sensitivity is $\approx$0.8\,K and $\approx$0.1\,K per channel in the high and low spectral resolution cubes, respectively.
	
\subsection{Spectra extraction}\label{sec:spectraextraction}

	An overview of the ALMA 3\,mm and JWST 5.2\,$\upmu$m and 19\,$\upmu$m continuum data is presented in Fig. \ref{fig:continuum}. The central core of IRAS\,18089 (labeled ``mm'') is embedded in an extended dusty envelope revealed by the 3\,mm continuum. There is no significant contribution from free-free emission at this wavelength \citep[less than 1\,mJy,][]{Zapata2006,Gieser2023a}. Extended emission is also detected in both MIR continuum maps which is in contrast to the JOYS high-mass target IRAS\,23385 that is also in the HMPO phase where only two point sources were detected at 5\,$\upmu$m \citep{Beuther2023, Gieser2023b}. The brightest source at 5.2\,$\upmu$m is located $\approx3''$ towards the south-west of the mm core, referred to as the MIR source in the following. Based on the presence of one Gaia source in the parallel 15\,$\upmu$m MIRI image, the astrometric correction would be less than 0.1$''$ and therefore no astrometric correction is applied \citep[Appendix B in ][]{ReyesReyes2026}. The offset between the mm and MIR peak is hence due to the presence of several sources where some are still highly embedded. We note the increase of the field-of-view in the MIRI-MRS data from short to long wavelengths in Fig. \ref{fig:continuum}.
	
	The 19\,$\upmu$m peak emission is slightly offset from both the mm and MIR peak position (0.9$''$ towards the south-east), which might be an effect of high extinction or the presence of more embedded sources. This position also corresponds to the emission peak at longer wavelengths (up to 28\,$\upmu$m) in the MIRI-MRS data. In this work, we focus on the chemical properties towards the position of the 3\,mm continuum peak position, which is the hot core region. Spectra are extracted from both the JWST and ALMA data at the position of the 3\,mm continuum peak at $\alpha$(J2000)=18$^{\mathrm{h}}$11$^{\mathrm{m}}$51.46$^{\mathrm{s}}$ and $\delta$(J2000)= $-$17$^\circ$31$'$28.75$''$. In both data sets, the aperture is set to a fixed radius of 1$''$ (red circle in Fig. \ref{fig:continuum}) covering a diameter of 5\,000\,au at the distance of IRAS\,18089 (2.5\,kpc). Fig. \ref{fig:continuum} also shows the position of the MIR peak position, $\alpha$(J2000)=18$^{\mathrm{h}}$11$^{\mathrm{m}}$51.40$^{\mathrm{s}}$ and $\delta$(J2000)= $-$17$^\circ$31$'$29.96$''$, analyzed in \citet{vanDishoeck2025}. In contrast to the hot core region, the MIR source is not associated with bright gas-phase molecular line emission in the ALMA data (Sect. \ref{sec:gas}). This MIR source is hence an ideal local background source for an ice composition comparison slightly offset from the hot core region.

	For each sub-band in the JWST data we apply a 1D residual fringe correction using \texttt{fit\_residual\_fringes\_1d} of the \texttt{jwst} python package (Kavanagh et al., in preparation). Following the method by \citet{vanGelder2024b}, individual spectra from the 12 sub-bands of the JWST/MIRI-MRS data are stitched to a full spectrum from 4.9 to 26$\upmu$m with small additive offsets being corrected for in each sub-band starting at the lowest wavelength data (ch1 short). We chose ch1 short as the reference since the photometric calibration is most accurate in this sub-band \citep{Argyriou2023}. The noise from ch1 to ch3 is $\approx$1\,mJy and increases to 5, 32, 150\,mJy in ch4 short, medium and long, respectively. Given the high MIR-brightness of IRAS\,18089, a high signal-to-noise ratio (S/N) ranging from 50 up to 850 is achieved.
	
	The observed MIR spectrum of IRAS\,18089 is presented in Fig. \ref{fig:spectrumIR}. The source spectrum has a large dynamic range covering four orders of magnitude. The lowest flux density is reached at $\approx$1\,mJy at $9-11$\,$\upmu$m due to strong silicate absorption which is broad and flat between 9 and 11\,$\upmu$m due to reaching the noise threshold. The SED then rises towards longer wavelengths to 10\,Jy at 26\,$\upmu$m, as expected from a young embedded protostar. The spectrum shows broad absorption features by molecular ices. The extraction of column densities of ice species is further explained in Sect. \ref{sec:ice}.
	
	In order to estimate molecular column densities of gas-phase species, we extract the mean spectrum within the same aperture size from all ALMA spws. The noise $\sigma_\mathrm{line}$ per channel, estimated in the 1$''$ radius extracted spectrum (Sect. \ref{sec:spectraextraction}), is $\approx$0.3\,K and $\approx$0.06\,K in the narrowband and broadband spws, respectively (Table \ref{tab:alma_obs}). In Sect. \ref{sec:gas} we derive molecular column densities detected within this data set.

\section{Results}\label{sec:results}

	The high-mass hot core IRAS\,18089 is expected to show a rich molecular composition both in the solid-state as well as the gas-phase. The MIR continuum SED of IRAS\,18089 and molecular constituents in the icy mantles around dust grains are analyzed in Sect. \ref{sec:ice} using sensitive MIRI-MRS observations obtained with JWST. Gas-phase molecules are characterized in Sect. \ref{sec:gas} using ALMA observations at 3\,mm.

\subsection{Ice column density determination using JWST MIRI-MRS spectra}\label{sec:ice}

	Figure \ref{fig:spectrumIR} presents the full MIRI-MRS spectrum of the IRAS\,18089 mm hot core exhibiting many distinct absorption features. The broad and deep absorption feature at 10\,$\upmu$m is due to silicates and the observed flat profile at the peak absorption is due to reaching the noise floor between $\approx$9\,$\upmu$m and 11\,$\upmu$m ($\approx$1\,mJy). In addition to silicates, broad absorption features due to molecular ices are present in the NIR and MIR range. Molecular column densities of such ice species can be quantified by fitting the ice optical depth spectrum $\tau_\mathrm{\nu}$ with reference laboratory spectra of pure ices as well as ice mixtures.
	
	To obtain the ice optical depth spectrum, the spectral energy distribution (SED) of the continuum has to be first determined from the MIRI-MRS data (Fig. \ref{fig:spectrumIR}). The continuum SED can be estimated, for example, by interpolation of guiding points set above the observed spectrum and with an additional absorption component of silicate dust $\tau_\mathrm{sil}$ \citep{Chen2024, Rocha2024, Rayalacheruvu2025}. Here, we use a physically-motivated approach \citep[similar to the MIR source analysis,][]{vanDishoeck2025} and model the continuum SED by considering black bodies $B_{\nu}(T)$, with $F_\nu$ = $\Omega B_{\nu}(T)$ and emitting area $\Omega$ and by taking into account absorption by carbonaceous and silicate dust as well as H$_{2}$O ice.
	 
	 The dust composition is a mixture of amorphous carbon and amorphous silicates \citep{Pollack1994}. In this work we consider for the silicate composition the most common contributors, namely olivine (MgFeSiO$_4$) and pyroxene (Mg$_{0.7}$Fe$_{0.3}$SiO$_{3}$):
	\begin{equation}
	\tau_\mathrm{dust}= \tau_\mathrm{olivine} + \tau_\mathrm{pyroxene} = f_\mathrm{oli}\times\kappa_\mathrm{oli} + f_\mathrm{pyr}\times\kappa_\mathrm{pyr}.
	\end{equation}	 	 
	We use reference opacity spectra ($\kappa_\mathrm{oli}$ and $\kappa_\mathrm{pyr}$ in cm$^{2}$\,g$^{-1}$) created with \texttt{OpTool} \citep{optool2021}, using a grain size distribution from 0.1 to 1\,$\upmu$m and taking into account amorphous carbon (with a volume fraction of 15\%), following the same approach as for the MIR source in IRAS\,18089 \citep[see Appendix F in][]{vanDishoeck2025}. The scaling factors $f_\mathrm{oli}$ and $f_\mathrm{pyr}$ measure the average mass surface density (in g\,cm$^{-2}$) of each dust component within the aperture.
	
	The absorption by water ice $\tau_\mathrm{H_2O}$ is also included in our SED model. This allows a direct fit of the SED model function to the observed spectrum between 10 and 15\,$\upmu$m (Fig. \ref{fig:spectrumIR}). Following the analysis approach of the HMSFR IRAS\,23385 MIRI-MRS spectrum \citep{Rocha2024}, the water ice composition is a combination of two temperature components at 15\,K and 75\,K:
	\begin{equation}
	\tau_\mathrm{H_2O} = f_\mathrm{1,H_2O}\times\tau_\mathrm{H_2O,15K}+f_\mathrm{2,H_2O}\times\tau_\mathrm{H_2O,75K} \mathrm{.}
	\end{equation}	
	Laboratory water ice optical depth spectra measured at 15\,K and 75\,K ($\tau_\mathrm{H_2O,15K}$ and $\tau_\mathrm{H_2O,75K}$) were taken from \citet{Oberg2007} with scaling factors $f_\mathrm{1,H_2O}$ and $f_\mathrm{2,H_2O}$, respectively, which consider different ratios of the two temperature components.
	
	To avoid over-fitting the spectrum with a model that contains too many free parameters, we iteratively started with one black body absorbed by dust and water which was not sufficient to explain the full complexity of the observed spectrum. Two black bodies with a hot and warm temperature component ($T_1$ and $T_2$) absorbed by dust and water were required to explain the full MIR spectrum of IRAS\,18089.
	
	We note that in reality, IRAS\,18089 has a more complex temperature structure, that is a temperature gradient as a function of radius \citep{Gieser2023a}. A full radiative transfer model, including the protostar, disk and envelope components, is beyond the scope of this work. Our SED model is sufficient to determine the continuum SED which is necessary to infer the H$_{2}$O ice column density that is overlapping with the silicate absorption feature at 10\,$\upmu$m (Fig. \ref{fig:spectrumIR}). The presence of the two black bodies at warm and hot temperatures can either indicate an underlying temperature gradient and/or two unresolved sources, which can also be expected in clustered HMSFRs. The complete SED model function, $F_{\nu,\mathrm{model}}$, is
	
	\begin{align*} 
	F_{\nu,\mathrm{model}} = &(\Omega_1 B_{\nu}(T_1) + \Omega_2 B_{\nu}(T_2)) \times e^{-( \tau_\mathrm{dust} + \tau_\mathrm{H_2O})} \\
	= &(\Omega_1 B_{\nu}(T_1) + \Omega_2 B_{\nu}(T_2)) \\
	&\times e^{-(f_\mathrm{oli}\times\kappa_\mathrm{oli} + f_\mathrm{pyr}\times\kappa_\mathrm{pyr} + f_\mathrm{1,H_2O}\times\tau_\mathrm{H_2O,15K}+f_\mathrm{2,H_2O}\times\tau_\mathrm{H_2O,75K})}\\
	\end{align*}
	
	and includes in total 8 free parameters, 4 for the two black body components and 4 for the scaling factors of the dust and water reference spectra. We use the \texttt{curve\_fit} function of the \texttt{scipy} python package \citep{scipy} and the \texttt{emcee} package \citep{ForemanMackey2013} to find the best fit to the observed spectrum and estimate the uncertainties using the Markov Chain Monte Carlo (MCMC) method. Given that the spectrum contains broad ice absorption features by species other than water ice and silicates (Fig. \ref{fig:spectrumIR}), we only select specific wavelength ranges outside these ice bands for the fit (marked in Fig. \ref{fig:spectrumIR}). The lower boundary for the dust and water fit parameters ($f_\mathrm{oli}$,$f_\mathrm{pyr}$,$f_\mathrm{1,H_2O}$,$f_\mathrm{2,H_2O}$) is set to 0, hence individual contributions can be 0 and are not forced to be present in the fit. Additional major ice constituents present in the MIR are labeled in Fig. \ref{fig:spectrumIR} \citep[based on][]{Whittet1996, Gibb2000, Yang2022,McClure2023}. In addition there is a broad feature at 11.3\,$\upmu$m caused by Polycyclic aromatic hydrocarbon (PAH) emission. There is a potential broad absorption feature at 16.3\,$\upmu$m that could be explained by the presence of crystalline silicates. A more detailed study of the dust composition towards IRAS\,18089 is however beyond the scope of this work.

\setlength{\tabcolsep}{4pt}
\begin{table}[!htb]
\caption{Best-fit SED model parameters.}
\label{tab:SEDmodel}
\centering
\renewcommand{\arraystretch}{1.4}
\begin{tabular}{ll}
\hline\hline
Parameter & Best fit value\\
 \hline\hline
Black body components & \\
$T_1$ (K) & 410$^{+17}_{-14}$\\
$\Omega_1$ (sr) & (8.0$^{+2.6}_{-1.9}$)$\times10^{-15}$\\
$T_2$ (K) & 83$^{+4.7}_{-4.0}$\\
$\Omega_2$ (sr) & (5.9$^{+3.2}_{-2.2}$)$\times10^{-11}$ \\
\hline
Dust composition & \\
$f_\mathrm{oli}$ (g\,cm$^{-2}$) & (2.3$^{+0.23}_{-0.23}$)$\times10^{-3}$ \\
$f_\mathrm{pyr}$ (g\,cm$^{-2}$) & (4.4$^{+2.2}_{-2.1}$)$\times10^{-4}$ \\
\hline
Water composition & \\
$f_\mathrm{1,H_2O}$ (15\,K) & 0.71$^{+1.0}_{-0.52}$ \\
$f_\mathrm{2,H_2O}$ (75\,K) & 8.4$^{+0.83}_{-1.3}$ \\
\hline
\end{tabular}
\end{table}

	The best-fit SED model, taking into account the continuum emission and absorption by dust and water, is shown in Fig. \ref{fig:spectrumIR} and the best-fit parameters are summarized in Table \ref{tab:SEDmodel}. The best-fit values of the MCMC SED fit are estimated from the $50^{th}$-percentile and the $-\sigma$ and $+\sigma$ uncertainties are computed from the difference to the $16^{th}$ and $84^{th}$-percentiles, respectively. Overall, there is a good agreement with the observed spectrum of IRAS\,18089. As expected from an embedded protostar, the black body components of the hot and warm component are $\approx$410\,K and 80\,K, respectively. These components are possibly tracing the inner dusty disk and the outer disk envelope. The hot temperature component of IRAS\,18089 mm is in agreement with the nearby brighter (at 5\,$\upmu$m) MIR source being slightly hotter \citep[$\approx$700\,K,][]{vanDishoeck2025}. The MCMC corner plot is presented in Fig. \ref{fig:MCMC_SED}, highlighting degeneracies between the temperature and emitting area of each black body component. There are also degeneracies between the two silicate components as well as the two temperature components of the H$_{2}$O ice.
	
	The best-fit water composition is a mixture of 15\,K and 75\,K indicating different temperature regimes along the line-of-sight. The total water column density, $N$(H$_{2}$O), is (1.5$\pm$0.75)$\times10^{19}$\,cm$^{-2}$ (see Eq. \ref{eq:icecoldens}). The column densities of the cold (15\,K) and warm (75\,K) water component contribute 12\% and 88\% to the total column density. For the high-mass binary IRAS\,23385, the contribution of the warm component was less \citep[only $\approx$30\%,][]{Rocha2024}. Full SED models of all other high-mass JOYS sources are presented in \citet{ReyesReyes2026} where overall the cold component dominates the modeled SEDs. Given the extended hot core region of the IRAS\,18089 mm source \citep{Gieser2023a}, a significant contribution of warm water ice is expected.

\begin{figure}[!htb]
\centering
\includegraphics[width=0.49\textwidth]{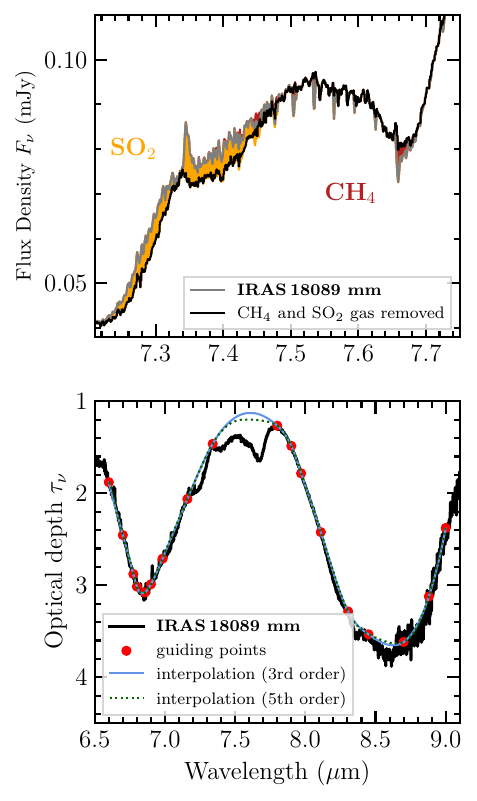}
\caption{Removal of gas-phase lines and local continuum estimate. The top panel shows the contribution of gas-phase SO$_{2}$ emission (yellow) and CH$_4$ absorption (red) lines estimated from the observed IRAS\,18089 mm spectrum (grey). The corrected spectrum is shown in black. The bottom panel presents the corrected optical depth spectrum (black) and the local continuum interpolation (blue line). The red dots mark the guiding points used for the interpolation. The green dashed line shows the continuum interpolation based on a 5th order polynomial (Appendix \ref{app:testsice}).}
\label{fig:spectrumtau}
\end{figure} 

\begin{figure*}[!htb]
\sidecaption
\centering
\includegraphics[width=0.7\textwidth]{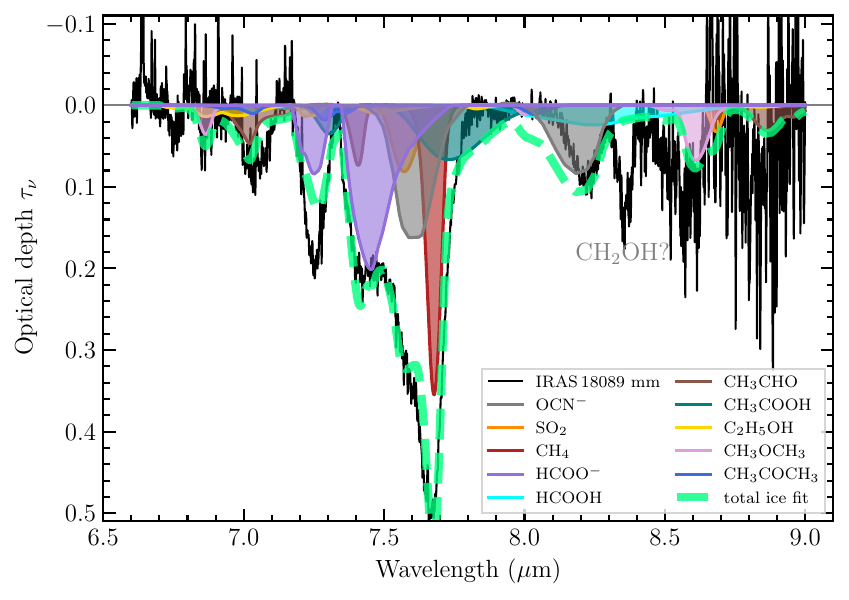}
\caption{Optical depth spectrum of IRAS\,18089 mm after local continuum subtraction. In black, the observed optical depth spectrum is presented highlighting the absorption features by minor ice constituents. The total fit considering a mixture of ice species is shown by the thick dashed green line. The solid lines are contributions by each ice species (details on the laboratory data are summarized in Table \ref{tab:ice_references}).}
\label{fig:spectrumICE}
\end{figure*} 

	The silicate absorption features at 10 and 17\,$\upmu$m are slightly overestimated in the SED model, due to reaching the noise limit in this regime in the MIRI-MRS data that is causing the flat profile at the bottom of the 10\,$\upmu$m silicate feature. A large portion of the blue part of the 10\,$\upmu$m silicate feature cannot be explained by the silicate composition of the best-fit model. We tried improving the fit by including an additional data point at 7.82\,$\upmu$m (Fig. \ref{fig:SEDmodel2}). While this does improve the fit at the blue part of the 10\,$\upmu$m silicate absorption, the temperature of the hot component is extreme ($T_1=1\,600$\,K). The resulting H$_{2}$O column density is 50\% less compared to the best-fit SED model (Table \ref{tab:SEDmodel}). Given the uncertainties in fitting the 10\,$\upmu$m silicate feature, we assume therefore that the H$_{2}$O ice column density is uncertain by 50\%.
	
		One explanation for the mismatch in 8-10\,$\upmu$m regime could be that this region contains a large amount of ice absorption and/or that the dust model is not accurate enough for a complex source such as the mm peak of IRAS\,18089. There could also be an additional contribution of silicate emission originating from the pseudo disk as observed towards irradiated disks in the HMSFR NGC\,6357 \citep{RamirezTannus2025}. While the observed spectrum towards the 10\,$\upmu$m silicate feature is affected by the noise floor, this is not the case for the 17\,$\upmu$m feature that is however also overestimated by our SED model. At these longer MIR wavelengths, the continuum emission is extended (Fig. \ref{fig:continuum}) and hence the environment emission contributes to the observed SED which is not taken into account in our SED model that only considers the source SED itself. This mismatch can also be explained by the dust composition towards the IRAS\,18089 hot core consisting of more than a simple mixture of amorphous carbonaceous material and silicates. Since the main focus of this work is to analyze the ice composition of COMs between 6.5\,$\upmu$m and 9\,$\upmu$m, our best-fit SED model is sufficient to obtain an estimate of the total water column density. But future work requires a more detailed investigation of the dust properties in MIRI-MRS spectra towards low- and high-mass protostars.
	
	The optical depth spectrum, taking into account the continuum SED with contributions by dust and water, is
	\begin{equation}
	\tau_\nu = -\mathrm{ln}\frac{F_{\nu\mathrm{,obs}}}{F_{\nu\mathrm{,model}}}
	\end{equation}
	with the observed MIRI-MRS spectrum $F_{\nu\mathrm{,obs}}$ and $F_{\nu\mathrm{,model}}$ obtained from the best-fit continuum SED model including absorption by dust and water (Table \ref{tab:SEDmodel}). The optical depth spectrum in the range from 6.5 to 9.0\,$\upmu$m is presented in the bottom panel in Fig. \ref{fig:spectrumtau}.

	To analyze the optical depth spectrum towards the range where many COM ices are present (``COM region'', $\approx$6.6-9.0\,$\upmu$m), a local continuum has to be first determined that removes contributions of the bulk ice features, i.e. the C-H stretch and OH-stretch at $\approx$6 and $\approx$7$\upmu$m (Fig. \ref{fig:spectrumIR}). The MIR spectrum of IRAS\,18089 shows not only deep ice absorption features, but also a plethora of gas-phase lines, including CH$_{4}$ in absorption and SO$_{2}$ in emission. These two species have strong transitions in the COM range from $\approx$7.2-7.7\,$\upmu$m. In particular, the presence of SO$_{2}$ gas-phase lines in emission create a pseudo-continuum that can cancel out ice absorption \citep{vanGelder2024a}.
	
	The CH$_{4}$ and SO$_{2}$ spectra are created using slab models assuming local thermal equilibrium (LTE) conditions \citep{vanGelder2024b,Francis2024}. The line width is assumed to be 4.7\,km\,s$^{-1}$ to be consistent with the slab modeling approach of the high-mass JOYS target IRAS\,23385 \citep{Francis2024}. This value is based on typical LTE models of T Tauri disks \citep{Salyk2011} and is in agreement with an average line width of $\approx$5\,km\,s$^{-1}$ measured for the gas-phase species with ALMA (Sect. \ref{sec:gas}). The gas temperature is fixed to 150\,K and for the SO$_{2}$ slab model we fix the emitting radius to 100\,au. We follow a similar approach as carried out for the ice analysis of the entire low-mass JOYS sample by Chen et al. (in preparation) to remove the contribution of CH$_{4}$ and SO$_{2}$ gas-phase lines in the spectrum. We manually vary the column density of each species until, by a visual inspection of the observed IRAS\,18089 spectrum, after subtracting the slab model spectra, the gas-phase line contributions are removed. We note that here we are not properly fitting the gas-phase species, so we refrain from further interpreting the results from the slab modeling as this would require a proper exploration of the parameter space (emitting radius, gas temperature, and column density). In the top panel in Fig. \ref{fig:spectrumtau} we present the comparison before and after gas-phase CH$_{4}$ and SO$_{2}$ removal. Both ice absorption bands at 7.3 and 7.4\,$\upmu$m become deeper after the correction. The importance of removing gas-phase contributions before analyzing the ices is further demonstrated in Appendix \ref{app:testsice}.
	
	 We then interpolate the local continuum in the optical depth spectrum towards the COM region using a set of guiding points and spline functions (Fig. \ref{fig:spectrumtau}). While in previous work the guiding points were often placed above the observed optical depth spectrum, here, we aim to stay as close as possible to the observed spectrum in order to extract only the most reliable absorption features of minor ice constituents. The resulting column densities might hence be slightly underestimated. \citet{Gross2026} analyzed the ices towards a sample of low-mass protostars using MIRI-MRS data and found that the choice of the local continuum baseline does affect the derived SO$_{2}$ ice column density which we further investigate in Appendix \ref{app:testsice} in the case of the IRAS\,18089 hot core. The final CH$_{4}$ and SO$_{2}$ gas-phase and local continuum subtracted optical depth spectrum is presented in Fig. \ref{fig:spectrumICE}. This spectrum reveals that within the broad and deep bulk ice absorption features, there is an additional contribution of minor ices, most notably the 7.7\,$\upmu$m absorption by CH$_{4}$.

	We follow the approach by \citet{Chen2024} to determine the ice column densities. We assume that the observed optical depth spectrum consists of a linear combination of different ice mixtures or pure ices taken from laboratory reference data $\tau_{\nu,m}^{\mathrm{lab}}$ scaled by a factor $\alpha_m$
	\begin{equation}
	\tau_\nu = \sum_{m}{\alpha_m\tau_{\nu,m}^{\mathrm{lab}}}.
	\label{eq:icefit}
	\end{equation}

	A list of all considered ices \citep[in total 12 different molecules motivated by JWST ice studies by][]{Rocha2024,Chen2024,Rayalacheruvu2025,Sewilo2025} is summarized in Table \ref{tab:ice_references}. This includes simple species (OCN$^{-}$, SO$_{2}$, CH$_{4}$) and COMs analyzed in previous studies \citep{Rocha2024,Chen2024,Rayalacheruvu2025,Sewilo2025}. Regarding the OCN$^{-}$ ice, in previous studies \citep{Rocha2024,Chen2024,Sewilo2025} only the 2$\nu_2$ band at 1295\,cm$^{-1}$ (7.72\,$\upmu$m), but not the $\nu_1$ band at 1210\,cm$^{-1}$ (8.26\,$\upmu$m) based on laboratory work by \citet{Novozamsky2001} and \citet{Raunier2003} was considered, whereas here we take into account both ice bands (Fig. \ref{fig:baseline_corr}). The strong $\nu_3$ band of OCN$^{-}$ at 4.64\,$\upmu$m (2155\,cm$^{-1}$) is not covered by the MIRI-MRS data and NIRSpec observations towards this source are currently not available.

	 Depending on the ice mixture composition, mixing ratio, and temperature, individual ice bands can change in position and shape. For many ice species, there are only limited experimental data available, with often H$_{2}$O as the bulk component measured at low temperatures ($\leq$15\,K). \citet{Chen2024} performed detailed tests on CH$_3$CHO and found that the mixture with H$_{2}$O at low temperature (15\,K) provides the best match between the laboratory work and observations (their Fig. 6). For most species, we use ice mixtures with water ice as the bulk component, if available in public databases, which is the most abundant ice component. In the case of SO$_{2}$ ice, only mixtures with CH$_{3}$OH were available. For H$_{2}$CO only pure ice spectra are available. For HCOOH it was easier to isolate HCOOH ice bands in the pure ice spectrum and similar in the case of CH$_3$OCHO using data from the CO mixture. Isolating the ice components from the bulk component(s) is important in our ice fitting approach to reliably estimate individual contributions. However future work requires measurements of ice mixtures containing many ice components, constrained by current JWST observations, which will allow a detailed analysis how the absorption features agree or disagree compared to simple mixtures. The laboratory spectra of ice mixtures contain contributions from, for example, H$_{2}$O or CH$_{3}$OH in the 6.5 to 9.0\,$\upmu$m wavelength range. In other cases the spectra are slightly offset from zero in ranges with no ice absorption. Hence we first apply a baseline correction to the laboratory spectra and to isolate the ice features of the minor ice constituents. Figure \ref{fig:baseline_corr} shows the laboratory spectra before and after baseline correction. The best-fit and uncertainties considering all ice constituents listed in Table \ref{tab:ice_references} are determined based on a least squares optimization using the \texttt{curve\_fit} package. In order to not bias the results, the lower boundary for the scaling factors $\alpha_m$ is 0, hence, in the fitting individuals ice constituents are not forced to be present.

	The ice column density is then calculated using the band strength $A$ of specific ice bands and integrating the best-fit scaled laboratory optical depth spectrum over the ice feature (in wave numbers $\tilde{\nu}$):
	\begin{equation}
	N_\mathrm{ice} = \frac{1}{A} \int_{\tilde{\nu_1}}^{\tilde{\nu_2}} \alpha \tau_{\tilde{\nu}}^{\mathrm{lab}} \mathrm{d}\tilde{\nu}.
	\label{eq:icecoldens}
	\end{equation}
	Table \ref{tab:ice_references} summarizes the considered molecules and their band strengths. For most ice bands (except for H$_{2}$O and OCN$^{-}$), we use the same band strength values as in \citet{Chen2024}. \citet{Mastrapa2009} measured band strengths for amorphous water ice within a range of $2.5\times10^{-17}$\,cm\,molecule$^{-1}$ to $3.0\times10^{-17}$\,cm\,molecule$^{-1}$ between 15\,K and 80\,K. For H$_{2}$O we use an average of $2.8\times10^{-17}$\,cm\,molecule$^{-1}$ \citep{Mastrapa2009}, compared to $3.2\times10^{-17}$\,cm\,molecule$^{-1}$ used in \citet{Chen2024} and \citet{Rocha2024}. \citet{Rocha2024} calculated the OCN$^{-}$ band strength of the 7.72\,$\upmu$m ice band ($A_{7.7}=7.5\times10^{-18}$\,cm\,molecule$^{-1}$) based on the 4.6\,$\upmu$m band \citep[$A_{4.6}=1.3\times10^{-16}$\,cm\,molecule$^{-1}$,][]{vanBroekhuizen2005}. Here, we use recent measurements on the 4.6\,$\upmu$m band strength \citep[$A_{4.6}=1.51\times10^{-16}$\,cm\,molecule$^{-1}$][]{Gerakines2025} and following the extrapolation by \citet{Rocha2024} estimate $A=8.7\times10^{-18}$\,cm\,molecule$^{-1}$ for the 7.72\,$\upmu$m ice band strength.

	The best-fit ice optical depth spectrum is presented in Fig. \ref{fig:spectrumICE}, where all individual ices are highlighted as well. Out of the 12 ice species included in the fitting, we find that 10 are present in the spectrum of IRAS\,18089 mm. We find no evidence for significant H$_{2}$CO and CH$_{3}$OCHO ice contributions to the observed spectrum with both scaling factors best-fitted with $\alpha_m=0$. Hence we refrain from reporting upper limits for those species. The ice column density uncertainties are calculated including the uncertainties obtained from the fit and an additional 20\% to take into account uncertainties of the local continuum determination and ice band strength. In Appendix \ref{app:testsice} we present additional tests of our ice fitting method. One caveat in this ice fitting approach is that all results are based on the assumption that no other ice constituents are present in the spectrum.
	
	The main constituents are CH$_{4}$, OCN$^{-}$ and HCOO$^{-}$ ices ($\tau_\mathrm{max}$>0.1). Minor contributors ($\tau_\mathrm{max}$ between 0.01 and 0.1) are SO$_{2}$, CH$_{3}$CHO, CH$_{3}$OCH$_{3}$, CH$_{3}$COOH, HCOOH, CH$_{3}$COCH$_{3}$, C$_{2}$H$_5$OH ices. The strong absorption feature at 8.35\,$\upmu$m that was also detected in the MIR source of IRAS\,18089 \citep{vanDishoeck2025} remains unidentified using public ice laboratory reference spectra. It is however likely originating from a CH$_{2}$OH band measured at 1197\,cm$^{-1}$ (corresponding to 8.35\,$\upmu$m) occurring when CH$_{3}$OH is exposed to UV irradiation \citep{Oberg2009,Yocum2021}. There is also an absorption feature of NH$_2$OH ice measured in laboratory experiments \citep[1191\,cm$^{-1}$,][]{Nightingale1954} and \citep[1201\,cm$^{-1}$,][]{Zheng2010} corresponding to the NH$_2$ rocking mode that is close to the unidentified observed absorption feature. However, the NOH bending mode measured at 1515\,cm$^{-1}$ (6.6\,$\upmu$m) and 1486\,cm$^{-1}$ (6.73\,$\upmu$m) by these experimental works, respectively, is not observed towards IRAS\,18089. Nevertheless we can not fully exclude the possibility that NH$_2$-group containing ice species could contribute to this feature. The absorption band at 7.2\,$\upmu$m can be only partially ($\approx$60\%) explained by a combination of HCOO$^{-}$, CH$_{3}$COOH, and C$_{2}$H$_5$OH ice. Small offsets of the best-fit compared to the observed spectrum are likely linked to the local continuum estimate (Fig. \ref{fig:spectrumtau}).
	
	The derived ice column densities, including H$_{2}$O from the SED model fit (Sect. \ref{sec:ice}), are summarized in Table \ref{tab:columdens} and range from $10^{16}$\,cm$^{-2}$ to $10^{19}$\,cm$^{-2}$. Given that the ice abundance of CH$_{3}$OH can not be inferred due to the flat plateau in the 10\,$\upmu$m silicate band (Fig. \ref{fig:spectrumIR}), the CH$_{3}$OH column density is inferred indirectly assuming the relative ice abundance $N$(CH$_{3}$OH)/$N$(H$_{2}$O) is the same for the mm and MIR peak position \citep[$\sim$0.18,][]{vanDishoeck2025}. For the H$_{2}$O and CH$_{3}$OH column density we assume that the uncertainties are 50\%. In Sect. \ref{sec:discussion} we compare the ice and gas abundances towards IRAS\,18089 mm and further compare the ice composition to other low- and high-mass protostars analyzed with recent JWST/MIRI-MRS data.

\subsection{Gas-phase abundances using mm spectra}\label{sec:gas}
\begin{figure*}[!htb]
\sidecaption
\centering
\includegraphics[width=0.7\textwidth]{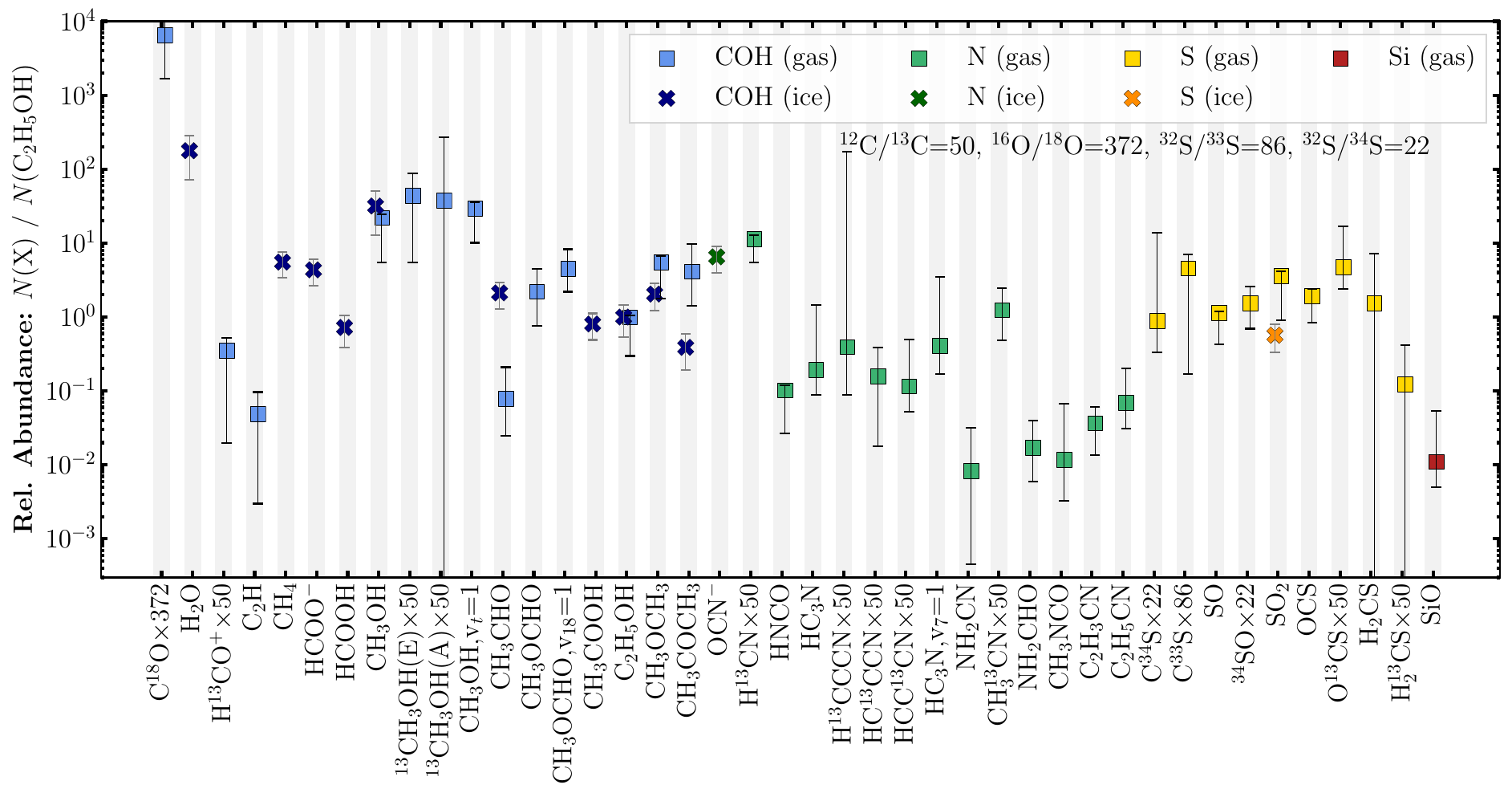}
\caption{Comparison of ice and gas-phase abundances (relative to C$_{2}$H$_{5}$OH) towards IRAS\,18089 mm. The crosses and squares mark ice and gas-phase abundances, respectively. Blue, green, yellow, and red colors highlight COH-, N-, S-, and Si-bearing species. The column density of less abundant isotopologues (including $^{13}$C, $^{18}$O, $^{33}$S, and $^{34}$S isotopes) were converted to their main isotopologue using relations by \citet{Wilson1994} and \citet{Yan2023} with a Galactocentric distance of $d_\mathrm{gal}$=5.7\,kpc for IRAS\,18089}.
\label{fig:comp_ice_gas}
\end{figure*} 
	Gas-phase molecules can be easily observed through their rotational levels at (sub)mm wavelengths. The observed 3\,mm spectra obtained with ALMA are presented in Figs. \ref{fig:alma} and \ref{fig:alma2}. Many spectral lines are detected covering simple molecular species with strong line emission, but there is in addition a forest of weak COM lines.
	
	Figure \ref{fig:ALMAline} shows line-integrated intensity maps of SiO, SO$_{2}$, HC$^{13}$CCN, $^{13}$CH$_{3}$OH, C$_{2}$H$_{5}$OH, and CH$_{3}$OCH$_{3}$ transitions. While SiO is tracing the outflow launched by the mm source \citep[consistent with low resolution SiO ($5-4$) data,][]{Beuther2004}, the spatially resolved emission of COMs, including less abundant isotopologues, as well as S- and N-bearing species are tracing the hot core region of IRAS\,18089 mm. Contrary to that, there is no significant molecular gas-phase emission at the location of the nearby MIR source.

	Molecular column densities of the gas-phase species are inferred using the \texttt{XCLASS} tool \citep{XCLASS}. With \texttt{XCLASS}, the radiative transfer equation of each identified molecule is solved assuming LTE. The optical depth $\tau_\nu$ of each transition is taken into account in the calculation of the synthetic model spectra, i.e. we do not assume that the transitions are all optically thin. The best-fit evaluated based on a least $\chi^2$ analysis. The fit parameters are the source size $\theta_\mathrm{source}$, rotation temperature $T_\mathrm{rot}$, column density $N$, FWHM line width $\Delta \varv$ and velocity offset $\varv_\mathrm{off}$. We fix the source size to the aperture size of the extracted spectra (diameter of 2$''$), hence we assume that the emission is spatially resolved (which is the case even for transitions of less abundant isotopologues, Fig. \ref{fig:ALMAline}) and that the beam filling factor is $\sim$1. 
	
	For the $\chi^2$ minimization procedure of the remaining four parameters we use an algorithm chain of the Genetic and Levenberg-Marquart algorithms using 50 iterations each per molecule. Uncertainties of the fit parameters are estimated using the MCMC algorithm \citep{ForemanMackey2013}. We fit in total 38 different species, also considering different isotopologues and torsionally or vibrationally excited states, all summarized in Table \ref{tab:columdens}. The ALMA 3\,mm spectra trace various types of molecules, including O-, N-, and S-bearing species as well as SiO. Table \ref{tab:columdens} also shows which catalog was used for each species for the \texttt{XCLASS} fitting. For each species, all spectral windows are included in the \texttt{XCLASS} fitting approach. Due to the line rich spectra, some transitions, in particular weaker COM lines, are blended at our spectral resolution of 0.9 and 3.4\,km\,s$^{-1}$ (Figs. \ref{fig:alma} and \ref{fig:alma2}). However since the data set covers a large bandwidth, the presence of many detected COM transitions ensures that line blending does not impact the \texttt{XCLASS} fit.
	
	The results for the gas column density of all species is listed in Table \ref{tab:columdens} and the total best-fit spectrum is presented in Figs. \ref{fig:alma} and \ref{fig:alma2}. In Sect. \ref{sec:discicegas} we compare and discuss the abundances of molecules in the gas-phase and in the ices.

\section{Discussion}\label{sec:discussion}

	In this work we analyzed both the molecular ice and gas-phase column densities towards the hot core region ($\sim$5\,000\,au) in IRAS\,18089. Sensitive JWST and ALMA observations reveal a rich ice chemical composition in the solid-state and gas-phase. In Sect. \ref{sec:discicegas} we discuss the ice and gas-phase abundances towards the IRAS\,18089 hot core in comparison with low-mass hot core sources \citep{Chen2024} and a hot core model \citep{Garrod2022}. In Sect. \ref{sec:discice} we further compare the ice composition of IRAS\,18089 with low- and high-mass star-forming regions targeted by recent JWST observations.

\subsection{Ice and gas-phase abundances in the IRAS\,18089 hot core}\label{sec:discicegas}

\begin{figure*}[!htb]
\sidecaption
\centering
\includegraphics[width=0.7\textwidth]{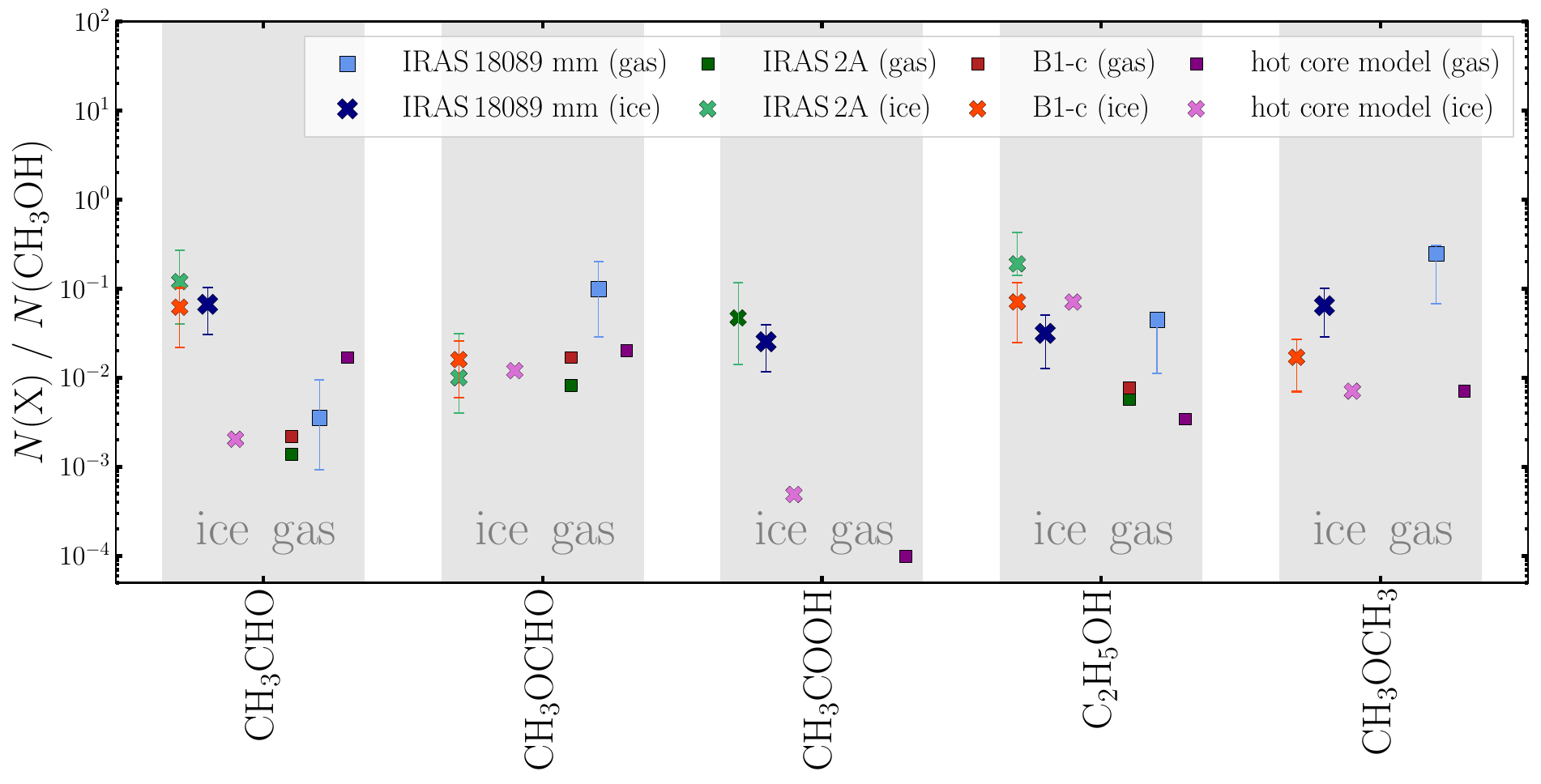}
\caption{Comparison of ice and gas-phase abundances (relative to CH$_{3}$OH) towards low- and high-mass hot cores. The crosses and squares mark ice and gas-phase abundances, respectively. Blue, green, and red data points are IRAS\,18089 mm (this work), IRAS\,2A \citep{Chen2024,Rocha2024}, and B1-c \citep{Chen2024}. The pink data points are taken from the hot core model by \citet{Garrod2022} (the fast warm-up model, at $T$=100\,K, their Fig. 10).}
\label{fig:comp_ice_lit_mod}
\end{figure*} 

	A comparison of relative ice and gas-phase abundances towards IRAS\,18089 is presented in Fig. \ref{fig:comp_ice_gas} using JWST/MIRI-MRS and ALMA 3\,mm observations (Table \ref{tab:columdens}). Given that gas-phase H$_{2}$O transitions were not covered in the ALMA 3\,mm setup and the CH$_{3}$OH ice column density could not be directly inferred from the JWST/MIRI-MRS data, we show molecular abundances relative to C$_{2}$H$_{5}$OH for which column density estimates are available for both gas and ice phases. Thus, in the following, our comparison is always based on abundances relative to C$_{2}$H$_{5}$OH, unless stated otherwise. For completeness, we show molecular abundances relative to CH$_{3}$OH (with the ice column density only indirectly inferred from the nearby MIR position, Sect. \ref{sec:ice}) in Fig. \ref{fig:comp_ice_gas_CH3OH}. With both C$_{2}$H$_{5}$OH and CH$_{3}$OH as reference species the same trends can be observed.
	
	In Fig. \ref{fig:comp_ice_gas} (as well as Fig. \ref{fig:comp_ice_gas_CH3OH} and Table \ref{tab:columdens}) the column density of less abundant isotopologues covered by ALMA were converted to their main isotopologue using $^{12}$C/$^{13}$C=50, $^{16}$O/$^{18}$O=372, $^{32}$S/$^{34}$S=22 \citep{Wilson1994} and $^{32}$S/$^{33}$S=86 \citep{Yan2023} with a Galactocentric distance of $d_\mathrm{gal}$=5.7\,kpc for IRAS\,18089. The results in Fig. \ref{fig:comp_ice_gas} are grouped by C/O-, N-, S- and Si-bearing species.

	In general, most O-bearing molecules have the highest relative abundance, followed by S-bearing, and then N-bearing molecules and with SiO having the lowest relative abundance. While CH$_{3}$OCHO is abundant in the gas-phase, we do not find significant contributions in the ice spectrum (Fig. \ref{fig:spectrumICE}). For HCOOH and CH$_{3}$COOH we carefully checked the ALMA spectra for line detections, and the non-detection in the gas-phase might be either linked to strong transitions not covered in the ALMA 3\,mm setup or these species being not abundant in the gas-phase. Assuming $T_\mathrm{rot}=100$\,K and $\Delta \varv=5$\,km\,s$^{-1}$ (mean values from all detected species), we estimate upper limits of $5\times10^{16}$\,cm$^{-2}$ and $1\times10^{16}$\,cm$^{-2}$ for the HCOOH and CH$_{3}$COOH gas-phase column densities which is less than the derived ice column densities (Table \ref{tab:columdens}).

\begin{figure*}[!htb]
\sidecaption
\centering
\includegraphics[width=0.7\textwidth]{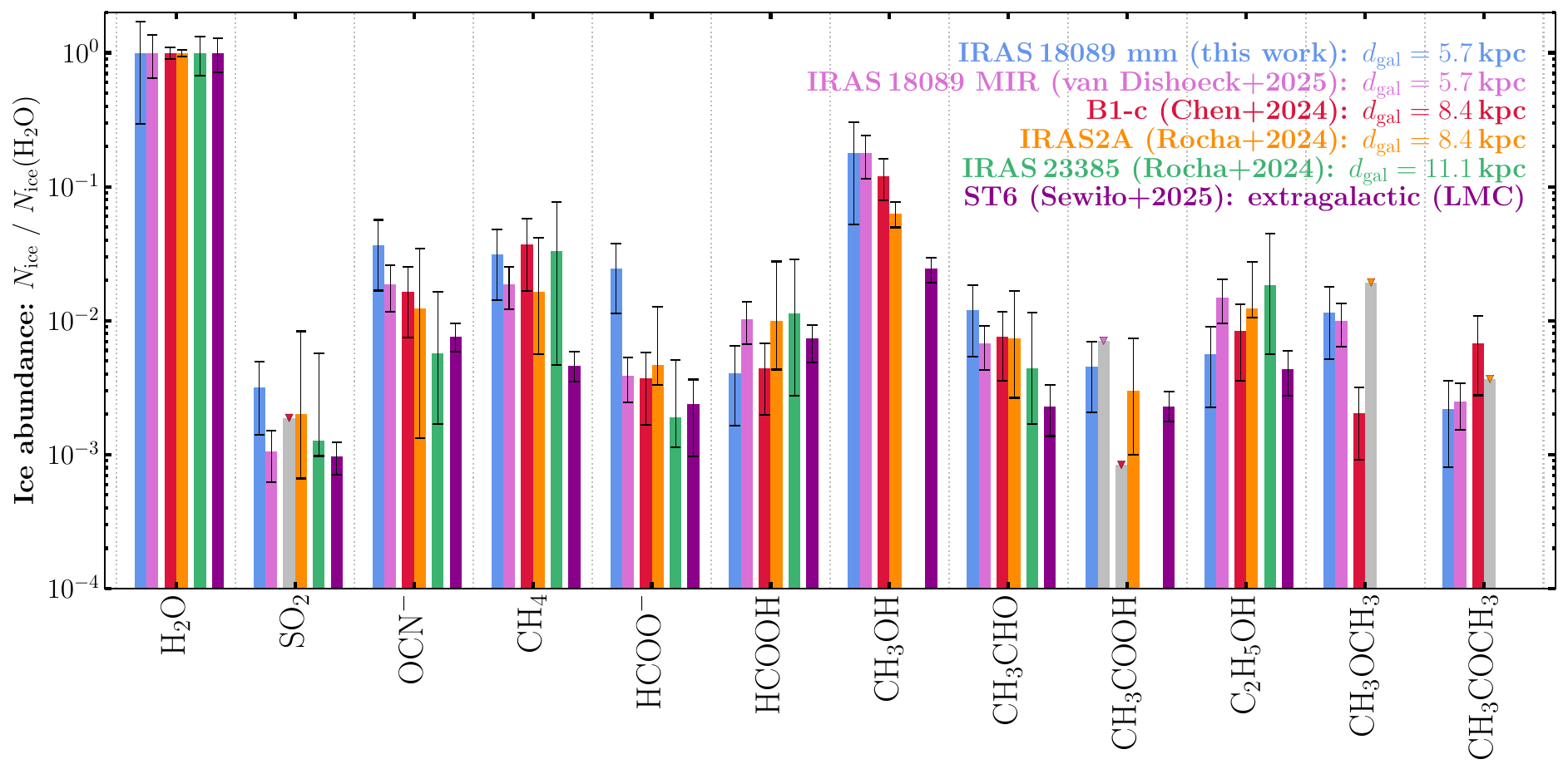}
\caption{Comparison of ice abundances (relative to H$_{2}$O) towards low- and high-mass protostars. The abundances towards the IRAS\,18089 mm analyzed in this work is shown in blue. The results for the low-mass protostars B1-c (red) and IRAS\,2A (orange) were taken from \citet{Chen2024}. The high-mass protostar IRAS\,23385 (green) and MIR peak of IRAS\,18089 (pink) were analyzed in \citet{Rocha2024} and \citet{vanDishoeck2025}, respectively. The ST6 source located in the LMC was studied by \citet{Sewilo2025} and is shown in purple. Upper limits are indicated by grey bars and triangle data points.}
\label{fig:comp_ice_lit}
\end{figure*}

	We find a good agreement between CH$_{3}$OH ice and gas-phase abundance, even though the CH$_{3}$OH ice column density could be only indirectly inferred from the IRAS\,18089 MIR source. This is consistent with the ice composition of IRAS\,18089 mm and MIR sources being similar in general (w.r.t. H$_{2}$O ice, Fig. \ref{fig:comp_ice_lit}), as further discussed in Sect. \ref{sec:discice}. Within the uncertainties we find a good agreement between CH$_{3}$OCH$_{3}$ ice and gas-phase abundances. Relative to CH$_{3}$OH, the C$_{2}$H$_{5}$OH abundance in the ice and gas-phase also agree with each other (Fig. \ref{fig:comp_ice_gas_CH3OH}). However, not all species show the same abundances in the gas and ice phase. For CH$_{3}$CHO, the abundance is more than one order of magnitude higher in the ice compared to the gas-phase and for SO$_{2}$ and CH$_{3}$COCH$_{3}$ the gas-phase abundance is about one order of magnitude higher compared to the ice abundance.

	We find a high abundance of OCN$^{-}$ ice, similar to HCN in the gas-phase. Recently, there has been a tentative detection of CH$_{3}$CN ice \citep{Nazari2024}. We checked the IRAS\,18089 ice spectrum (Fig. \ref{fig:spectrumICE}), but similar to other low- and high-mass protostars \citep{Chen2024, vanDishoeck2025, Sewilo2025}, there is no detection of CH$_{3}$CN in the ice using MIRI-MRS data. A more detailed investigation of N-bearing ices requires additional NIRSpec observations \citep{Nazari2024}.
	
	Many S-bearing isotopologues (from CS, SO, SO$_{2}$, OCS and H$_{2}$CS) are very abundant in the gas-phase. Our results for the SO$_{2}$ ice and gas-phase abundance relative to CH$_{3}$OH (0.02 and 0.1, respectively) are in agreement with a recent study by \citet[][their Fig. 6]{Santos2024b}. These authors compared the gas-phase SO$_{2}$ abundance (relative to CH$_{3}$OH) obtained in a sample of HMSFRs with ALMA with ice abundances from \citet{Boogert1997} and obtain $\approx0.1$ for both phases. Only one detection and one upper limit could be inferred for SO$_2$ ice towards HMSFRs in the study by \citet{Boogert1997} highlighting the need for more measurements in the future. There is a hint at OCS ice absorption at 4.9\,$\upmu$m in the MIRI-MRS spectrum (Fig. \ref{fig:spectrumIR}), however it is at the edge of the covered wavelength range. OCS ice has been detected towards many HMSFRs \citep{Boogert2022}. Additional JWST/NIRSpec data is essential to better constrain the sulfur inventory of the ices. Since the simpler S-bearing molecule ice features are covered by NIRSpec, for which IRAS\,18089 has no available data, and only a few laboratory works were dedicated to S-bearing COMs \citep{Santos2024,Narayanan2025}, future effort towards S-bearing species is necessary.
	
	\citet{Chen2024} has conducted a direct comparison between ice and gas-phase abundances of COMs towards the two low-mass protostars IRAS2A and B1-c \citep[which are both low-mass hot core sources located in the Perseus molecular cloud, e.g.,][]{Bottinelli2007, vanGelder2022, Busch2025}. The ice composition of IRAS2A was previously analyzed by \citet{Rocha2024}. In Fig. \ref{fig:comp_ice_lit_mod} we show ice and gas-phase abundances of COMs (relative to CH$_{3}$OH) detected towards IRAS\,18089 mm and include the results obtained for the low-mass hot cores studied by \citet{Chen2024} (excluding data for which only upper limits could be constrained).
	
	A comprehensive gas-grain chemical model of a hot core has been carried out by \citet{Garrod2022}. One of their model setups included a fast warm-up stage, which is most suitable for high-mass hot cores that evolve on fast evolutionary timescales. For comparison we show in Fig. \ref{fig:comp_ice_lit_mod} the ice and gas-phase abundances relative to CH$_{3}$OH for this model setup at $T=100$\,K (their Fig. 10).
	
	Figure \ref{fig:comp_ice_lit_mod} reveals that the low- and high-mass hot cores have comparable (within one order of magnitude) ice abundances (relative to CH$_{3}$OH) for all presented species. This highlights that for COM formation the early cold stage is essential. The gas-phase abundances of CH$_3$OCHO and C$_{2}$H$_{5}$OH are enhanced in the case of the high-mass hot core. The ice abundances of the hot core model agree with the observed hot cores for CH$_{3}$OCHO, C$_{2}$H$_{5}$OH, and CH$_{3}$OCH$_{3}$. The ice abundances of CH$_{3}$CHO and CH$_{3}$COOH are under-predicted in the hot core model by more than one order of magnitude.
	
	In the case of CH$_{3}$CHO, the hot core model predicts in addition a higher gas-phase abundance compared to what is observed for the hot cores. Despite taking into account more reaction pathways for the formation of CH$_{3}$CHO in the hot core model \citep[Sect. 5.3 in][]{Garrod2022}, whereas only CH$_{3}$ + HCO was considered the dominant reaction in diffusion-only models, there is a systematic difference between observations and models. 
	
	We find that for the IRAS\,18089 hot core region, some species have similar abundances in the ice and gas-phase (CH$_{3}$OH, C$_{2}$H$_{5}$OH, CH$_{3}$OCH$_{3}$). On the other hand, SO$_{2}$ and CH$_{3}$COCH$_{3}$ have higher abundances in the gas-phase, suggesting the importance of additional gas-phase routes. Most notably, in the case of CH$_{3}$CHO, we find a disagreement of the predicted ice and gas-phase column density by the models from \citet{Garrod2022} in contrast to the observed values towards low-mass \citep{Chen2024} and high-mass hot cores (this work). The hot core model by \citet{Garrod2022} predicts that the CH$_{3}$CHO ice abundance should be one order of magnitude lower compared to its gas-phase abundance. However, the observations towards both low-mass and high-mass hot cores reveal an opposite trend with the CH$_{3}$CHO ice abundance 1-2 order of magnitudes higher than what is measured in the gas-phase. Future work requires a statistical analysis of ice and gas abundances of low- and high-mass protostars as well as applying physical-chemical modeling of specific sources to better understand the chemical links between different molecules \citep{Simons2020, Gieser2021,Garrod2022}.

	\subsection{Ice composition in low- and high-mass protostars}\label{sec:discice}

	Spectroscopic observations with JWST allow a detailed characterization of minor ice constituents. Due to the overlap of ice bands of many different molecular species in the 6.5 to 9.0\,$\upmu$m range (Fig. \ref{fig:spectrumICE}) the extraction of individual components is non trivial, and thus the ice composition in this range has so far only been analyzed in a limited sample of protostars observed with JWST. The composition can be determined by using a linear combination of different molecular ices \citep[][and this work]{Chen2024} or using tools such as \texttt{ENIIGMA} \citep{Rocha2021,Rocha2024} and \texttt{INDRA} \citep{Rayalacheruvu2025}. In this section we focus on a comparison between ice abundances (relative to H$_{2}$O ice) of the protostars for which their ice composition has been analyzed so far. We note that even though different methods were used to estimate ice column densities, overall these different techniques provide similar results.
	
	A comparison of ice abundances (relative to H$_{2}$O ice) for in total five protostars with the chemical inventory of the ices of the IRAS\,18089 mm hot core is presented in Fig. \ref{fig:comp_ice_lit}. The ices towards the low-mass protostars B1-c and IRAS\,2A were studied in \citet{Chen2024} and \citet{Rocha2024} and towards the high-mass protostars IRAS\,23385 and the MIR source of IRAS\,18089 in \citet{Rocha2024} and \citet{vanDishoeck2025}, respectively. We also compare our results with the ice composition of the ST6 protostar located in the Large Magellanic Cloud (LMC) which is a lower metallicity environment compared to the Milky Way \citep{Sewilo2025}. The Galactic sources in Fig. \ref{fig:comp_ice_lit} are sorted by Galactocentric distance.
	
	In general, SO$_{2}$ has a similar relative ice abundance of $\approx$10$^{-3}$. For OCN$^-$ ice we find a slightly lower ice abundances towards low metallicity environments, but note that in previous work a second ice band was not considered (Sect. \ref{sec:ice}). For HCOO$^-$ we find an ice abundance about one order of magnitude higher compared to other protostars. We note that the location of the two HCOO$^-$ ice bands (7.2 and 7.4\,$\upmu$m) are sensitive to the local continuum estimate (Fig. \ref{fig:spectrumtau}). However, the high abundance can be explained by thermal processing due to protostellar heating. In laboratory experiments, the amount of HCOO$^{-}$ ions is increased after heating ice mixtures containing NH$_{3}$ and HCOOH \citep[Sect. 4 in][]{Galvez2010}.
	
	For most of the analyzed ice species, the ice abundances are within the uncertainties similar across the different Galactic protostars, namely SO$_{2}$, CH$_4$, HCOOH, CH$_{3}$OH, CH$_{3}$CHO, CH$_{3}$COOH, C$_{2}$H$_5$OH, and CH$_{3}$COCH$_{3}$. For CH$_{3}$OCH$_{3}$ the ice abundance of both IRAS\,18089 mm and MIR are about one order of magnitude higher compared to the low-mass hot core B1-c. For the extragalactic ST6 source the ice abundances are systematically lower. This can be linked to the lower metallicity in the LMC compared to the Milky Way. In addition, there are hints at trends with Galactocentric radius for OCN$^{-}$, CH$_{3}$OH, and CH$_{3}$CHO. Due to a decreasing metallicity with increasing Galactocentric distance, the destruction of molecules is more efficient.

	In summary, within the uncertainties we do not find large variations of ice abundances among different Galactic sources. This stresses the importance of grain-surface chemistry in the cold stages of the star formation phase. Comparing the ices with the ST6 source in the LMC highlights the impact of metallicity on the molecular composition. Further studies at different Galactic radii are essential in order to investigate this trend within the Milky Way. However, since the number of analyzed sources (in both their gas-phase and ice composition) is still low, two low-mass protostars \citep{Chen2024,Rocha2024} and one high-mass protostar (this work), to investigate general trends requires a larger sample. However our results underline the importance of sensitive spectroscopic observations with JWST (including the recently approved programs with IDs 5804 and 8887) to better understand the formation of molecules in the interstellar medium.
		
\section{Conclusions}\label{sec:conclusions}

	In this work we analyzed the molecular composition in the ice and gas-phase towards the hot core IRAS\,18089 mm using sensitive and high resolution observations with JWST/MIRI-MRS and ALMA. Our main findings are summarized in the following.

	\begin{enumerate}
	\item The JWST/MIRI-MRS spectrum from 5 to 28\,$\upmu$m shows emission and absorption lines by gas-phase species as well as absorption features by silicates and ices. The continuum SED is well characterized by two embedded (hot $T$=410\,K, and warm, $T$=80\,K) modified black bodies, absorbed by carbonaceous and silicate dust and water ice. The total H$_{2}$O water ice column density is 1.5$\times10^{19}$\,cm$^{-2}$. Due to the strong silicate absorption at 10\,$\upmu$m reaching the noise floor, a behavior expected for other bright high-mass protostars as well, the H$_{2}$O water ice column density can only be constrained within an uncertainty of 50\% and also prevents a direct estimate of the CH$_{3}$OH ice column density.
	\item The optical depth spectrum towards the ``COM region'' (from 6.5 to 9.0\,$\upmu$m) consists of a variety of different ice species including SO$_{2}$, OCN$^{-}$, CH$_{4}$, HCOO$^{-}$, HCOOH, CH$_{3}$CHO, CH$_{3}$COOH, C$_{2}$H$_5$OH, CH$_{3}$OCH$_{3}$, and CH$_{3}$OCH$_{3}$. The column density of these ices are in the range of $\sim10^{16}-10^{17}$\,cm$^{-2}$.
	\item The ALMA data at 3\,mm reveal line-rich spectra originating from a variety of different gas-phase molecules from simple diatomic species to COMs, including O-, N-, S-bearing species as well as less abundant isotopologues and torsionally or vibrationally excited transitions. On average, we find that gas-phase abundances of O-bearing COMs are about 1 to 2 orders of magnitude higher compared to S-bearing and N-bearing species, respectively.
	\item Comparing the IRAS\,18089 mm ice and gas-phase abundances (relative to C$_{2}$H$_5$OH or CH$_{3}$OH), we find similar abundances in both phases for C$_{2}$H$_5$OH, CH$_{3}$OH, CH$_{3}$OCH$_{3}$. The abundances of SO$_{2}$ and CH$_{3}$COCH$_{3}$ are elevated in the gas-phase suggesting additional gas-phase formation routes. The ice abundance of CH$_{3}$CHO is one order of magnitude higher in the ices compared to the gas-phase, opposite of what is predicted from a hot core chemical model by \citet{Garrod2022}. Overall we find the same trends for this high-mass hot core studied in this work and two low-mass hot cores analyzed by \citet{Chen2024}.
	\item The ice abundances (relative to H$_{2}$O ice) towards IRAS\,18089 mm are comparable to other Galactic low- and high-mass protostars. In particular, the ice abundances are similar to the nearby IRAS\,18089 MIR position providing a nearby background source to probe the ice composition offset from the hot core region. There are hints of a decreasing abundance with Galactocentric distance for OCN$^{-}$, CH$_{3}$OH, and CH$_{3}$CHO.
	\end{enumerate}
	 
	JWST/MIRI-MRS observations have revealed a complex composition of molecular ices towards a few low- and high-mass star-forming regions. To better understand the variations of the ice and gas compositions between regions, a larger sample across different environments and evolutionary stages has to be probed. To better characterize N-bearing and S-bearing ices, additional NIRSpec data of high-mass protostars are essential.

\begin{acknowledgements}
	This work is based on observations made with the NASA/ESA/CSA James Webb Space Telescope. The data were obtained from the Mikulski Archive for Space Telescopes at the Space Telescope Science Institute, which is operated by the Association of Universities for Research in Astronomy, Inc., under NASA contract NAS 5-03127 for JWST. These observations are associated with program \#1290. The following National and International Funding Agencies funded and supported the MIRI development: NASA; ESA; Belgian Science Policy Office (BELSPO); Centre Nationale d’Etudes Spatiales (CNES); Danish National Space Centre; Deutsches Zentrum fur Luft und Raumfahrt (DLR); Enterprise Ireland; Ministerio De Economi\'a y Competividad; Netherlands Research School for Astronomy (NOVA); Netherlands Organisation for Scientific Research (NWO); Science and Technology Facilities Council; Swiss Space Office; Swedish National Space Agency; and UK Space Agency. This paper makes use of the following ALMA data: ADS/JAO.ALMA\#2018.1.00424.S ALMA is a partnership of ESO (representing its member states), NSF (USA) and NINS (Japan), together with NRC (Canada), NSTC and ASIAA (Taiwan), and KASI (Republic of Korea), in cooperation with the Republic of Chile. The Joint ALMA Observatory is operated by ESO, AUI/NRAO and NAOJ.
	
	A.C.G. acknowledges support from PRIN-MUR 2022 20228JPA3A “The path to star and planet formation in the JWST era (PATH)” funded by NextGeneration EU and by INAF-GoG 2022 “NIR-dark Accretion Outbursts in Massive Young stellar objects (NAOMY)” and Large Gran INAF-2024 “Spectral Key fea-tures of Young stellar objects: Wind-Accretion LinKs Explored in the infraRed (SKYWALKER)”.
	
	JMV acknowledges support from the Academy of Finland grant No 348342.

	Astrochemistry in Leiden is supported by funding from the European Research Council (ERC) under the European Union’s Horizon 2020 research and innovation program (grant agreement No. 291141 MOLDISK), and by NOVA and NWO through TOP-1 grant 614.001.751 and its Dutch Astrochemistry Program (DANII). The present work is closely connected to ongoing research within InterCat, the Center for Interstellar Catalysis located in Aarhus, Denmark.
\end{acknowledgements}

\bibliographystyle{aa}
\bibliography{bibliography}

\begin{appendix}

\section{ALMA observations}

	Molecular column densities of gas-phase species are derived using ALMA observations at 3\,mm wavelengths. The observations are described in Sect. \ref{sec:ALMAobs} and more details of the data calibration and imaging procedure are presented in \citet{Gieser2023a}. Table \ref{tab:alma_obs} is a summary of the observational parameters (frequency coverage, spectral resolution, synthesized beam) of all 23 spws that were observed in two separate spectral setups (referred to as SPR1 and SPR2). In each spectrum, the noise was estimated from the extracted spectrum towards the IRAS\,18089 mm peak position within an aperture radius of 1$''$ (Sect. \ref{sec:spectraextraction}).

\begin{table*}[!htb]
\caption{Overview of the ALMA spectral line observations at 3\,mm.}
\label{tab:alma_obs}
\centering
\renewcommand{\arraystretch}{1.3}
\begin{tabular}{lrrrrr}
\hline\hline
Spectral window & \multicolumn{3}{c}{Spectral setup} & Synthesized beam & Noise\\
 & $\nu_\mathrm{min}$ & $\nu_\mathrm{max}$ & $\delta\nu$ & $\theta_\mathrm{maj}\times\theta_\mathrm{min}$ (PA) & $\sigma_\mathrm{line}$\\
 & GHz & GHz & MHz & $''\times''$ ($^\circ$) & K \\
\hline
SPR1 spw1 & 86.624 & 86.732 & 0.244 & 0.77$\times$0.72 (91) & 0.27\\
SPR1 spw2 & 86.708 & 86.815 & 0.244 & 0.77$\times$0.71 (91) & 0.29\\
SPR1 spw3 & 86.801 & 86.908 & 0.244 & 0.77$\times$0.71 (91) & 0.28\\
SPR1 spw4 & 86.293 & 86.401 & 0.244 & 0.77$\times$0.72 (92) & 0.31\\
SPR1 spw5 & 87.389 & 87.496 & 0.244 & 0.77$\times$0.72 (92) & 0.27\\
SPR1 spw6 & 88.585 & 88.693 & 0.244 & 0.75$\times$0.71 (84) & 0.32\\
SPR1 spw7 & 87.879 & 87.986 & 0.244 & 0.77$\times$0.71 (89) & 0.25\\
SPR1 spw8 & 87.271 & 87.378 & 0.244 & 0.77$\times$0.72 (92) & 0.31\\
SPR1 spw9 & 97.67 & 97.777 & 0.244 & 0.7$\times$0.65 (77) & 0.25\\
SPR1 spw10 & 97.935 & 98.043 & 0.244 & 0.7$\times$0.65 (78) & 0.27\\
SPR1 spw11 & 96.874 & 96.981 & 0.244 & 0.69$\times$0.65 (80) & 0.34\\
SPR1 spw12 & 96.942 & 97.05 & 0.244 & 0.71$\times$0.66 (78) & 0.25\\
SPR1 spw13 & 99.28 & 99.387 & 0.244 & 0.69$\times$0.65 (77) & 0.3\\
SPR1 spw14 & 100.686 & 100.793 & 0.244 & 0.69$\times$0.64 (79) & 0.29\\
SPR1 spw15 & 99.254 & 99.362 & 0.244 & 0.69$\times$0.65 (77) & 0.3\\
SPR1 spw16 & 99.821 & 99.928 & 0.244 & 0.69$\times$0.65 (77) & 0.28\\
\hline
SPR2 spw1 & 108.731 & 108.839 & 0.244 & 1.75$\times$1.0 (106) & 0.14\\
SPR2 spw2 & 110.153 & 110.261 & 0.244 & 1.73$\times$0.99 (106) & 0.23\\
SPR2 spw3 & 109.415 & 109.522 & 0.244 & 1.74$\times$1.0 (106) & 0.15\\
SPR2 spw4 & 109.733 & 109.841 & 0.244 & 1.74$\times$1.0 (106) & 0.18\\
SPR2 spw5 & 108.607 & 110.325 & 0.977 & 1.69$\times$0.93 (106) & 0.07\\
SPR2 spw6 & 96.106 & 97.824 & 0.977 & 1.91$\times$1.05 (105) & 0.06\\
SPR2 spw7 & 97.906 & 99.624 & 0.977 & 1.87$\times$1.03 (106) & 0.06\\
\hline
\end{tabular}
\tablefoot{A detailed description of the observations and data reduction is presented in \citet{Gieser2023a}.}
\end{table*}

\section{SED model uncertainty estimate}

\begin{figure*}[!htb]
\sidecaption
\centering
\includegraphics[width=0.7\textwidth]{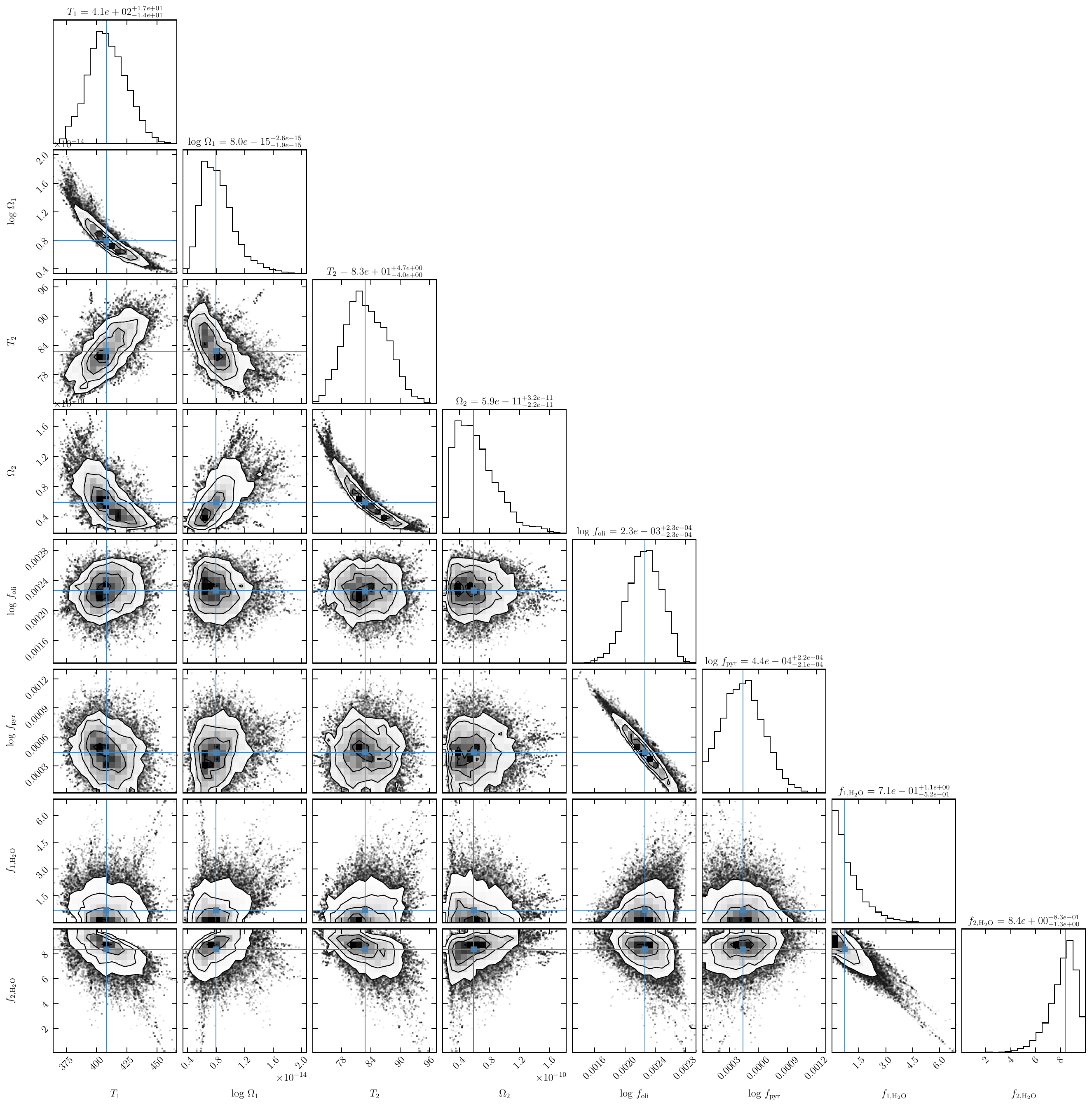}
\caption{Uncertainty estimate of the IRAS\,18089 SED model fit (Fig. \ref{fig:spectrumIR}) using the \texttt{emcee} package.}
\label{fig:MCMC_SED}
\end{figure*}

\begin{figure*}[!htb]
\sidecaption
\centering
\includegraphics[width=0.7\textwidth]{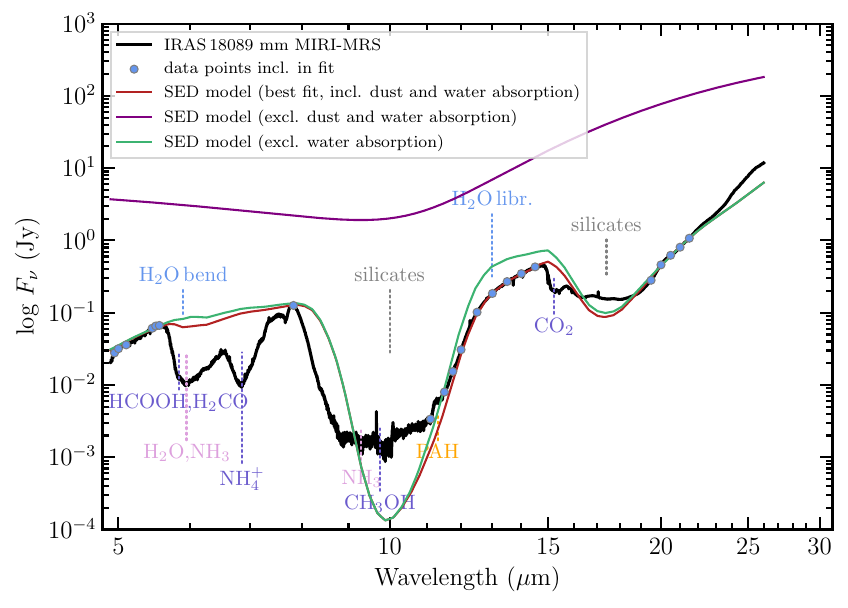}
\caption{The same as Fig. \ref{fig:spectrumIR}, but including the 7.82\,$\upmu$m data point in the fit.}
\label{fig:SEDmodel2}
\end{figure*}

	The uncertainties of the SED model (Sect. \ref{sec:ice}) are estimated using the \texttt{emcee} package \citep{ForemanMackey2013}. The corner plot is presented in Fig. \ref{fig:MCMC_SED}. Fig. \ref{fig:SEDmodel2} shows the SED fit when an additional data point at 7.82\,$\upmu$m is included in the fit that results in a better fit at the blue side of the 10\,$\upmu$m silicate absorption (compared to Fig. \ref{fig:spectrumIR}), but in a high temperature of the hot black body component ($T>1\,000$\,K).

\section{Reference laboratory spectra of molecular ices}\label{app:baselinecorr}

\setlength{\tabcolsep}{1pt}
\begin{table*}[!htb]
\caption{Laboratory data of ices and ice mixtures.}
\label{tab:ice_references}
\centering
\renewcommand{\arraystretch}{1.1}
\begin{tabular}{lrlr|lrl}
\hline\hline
Ice & & Mode & Band strength & Laboratory data & & Reference \\
& $\lambda$ & & $A$ $^{(a)}$ & & $T$ & \\
& ($\upmu$m) & & (cm\,molec.$^{-1}$) & & (K) & \\
\hline
H$_{2}$O & 13 & Libration & 2.8($-$17) & pure H$_{2}$O & 15 & \citet{Oberg2007}\\
& & & & & 75 & \citet{Oberg2007}\\ \hline
OCN$^{-}$ & 7.6 & Comb. (2$\nu_2$) & 8.7($-$18) & OCN$^-$:NH$_4^+$ & 80 & \citet{Novozamsky2001} \\
SO$_{2}$ & 7.6 & SO$_{2}$ stretch & 3.4($-$17) & SO$_{2}$:CH$_{3}$OH (1:1) & 10 & \citet{Boogert1997}\\
H$_{2}$CO & 8.0 & CH$_{2}$ rock & 1.5($-$18) & pure H$_{2}$CO & 15 & \citet{TerwisschavanScheltinga2021}\\
CH$_4$ & 7.7 & CH$_4$ deform & 8.4($-$18) & CH$_4$:H$_{2}$O (1:10) & 16 & \citet{Rocha2017}\\
HCOO$^{-}$ & 7.4 & CO\,stretch & 1.7($-$17) & NH$_4$COOH:H$_{2}$O (7:100) & 14 & \citet{Galvez2010} \\
HCOOH & 8.2 & C$-$O stretch & 2.9($-$17) & pure HCOOH & 15 & \citet{Bisschop2007}\\
CH$_{3}$CHO & 7.4 & CH$_{3}$ sym. def.+CH wag & 4.1($-$18) & CH$_{3}$CHO:H$_{2}$O (1:20) & 15 & \citet{TerwisschavanScheltinga2018} \\
CH$_{3}$OCHO & 8.3 & C$-$O stretch & 2.5($-$17) & CH$_{3}$OCHO:CO (1:20) & 15 & \citet{TerwisschavanScheltinga2021} \\
CH$_{3}$COOH & 7.8 & OH bend & 4.6($-$17) & CH$_{3}$COOH:H$_{2}$O (1:10) & 16 & $^{(b)}$\\
C$_{2}$H$_5$OH & 7.2 & CH$_{3}$ sym. def. & 2.4($-$18) & C$_{2}$H$_5$OH:H$_{2}$O (1:20) & 15 & \citet{TerwisschavanScheltinga2018} \\
CH$_{3}$OCH$_{3}$ & 8.6 & COC stretch+CH$_{3}$ rock & 5.6($-$18) & CH$_{3}$OCH$_{3}$:H$_{2}$O (1:20) & 15 & \citet{TerwisschavanScheltinga2018} \\
CH$_{3}$COCH$_{3}$ & 7.3 & CH$_{3}$ asym. stretch & 1.0($-$17) & CH$_{3}$COCH$_{3}$:H$_{2}$O (1:20) & 15 & \citet{Rachid2020}\\
\hline
\end{tabular}
\tablefoot{The format of the band strength is $a$($b$)=$a\times10^{b}$. All data, except for CH$_{3}$COOH, were taken from the The Leiden Ice Database for Astrochemistry \citep{Rocha2022}. \\ $^{(a)}$ Band strengths are taken from \citet[][and references within]{Chen2024}, except for H$_{2}$O \citep{Mastrapa2009} and OCN$^{-}$ \citep{Rocha2024,Gerakines2025}.\\$^{(b)}$: M. H. Moore et al. at NASA's Goddard Space Flight Center (\url{https://science.gsfc.nasa.gov/691/cosmicice/spectra.html})}
\end{table*}

\begin{figure}[!htb]
\centering
\includegraphics[width=0.49\textwidth]{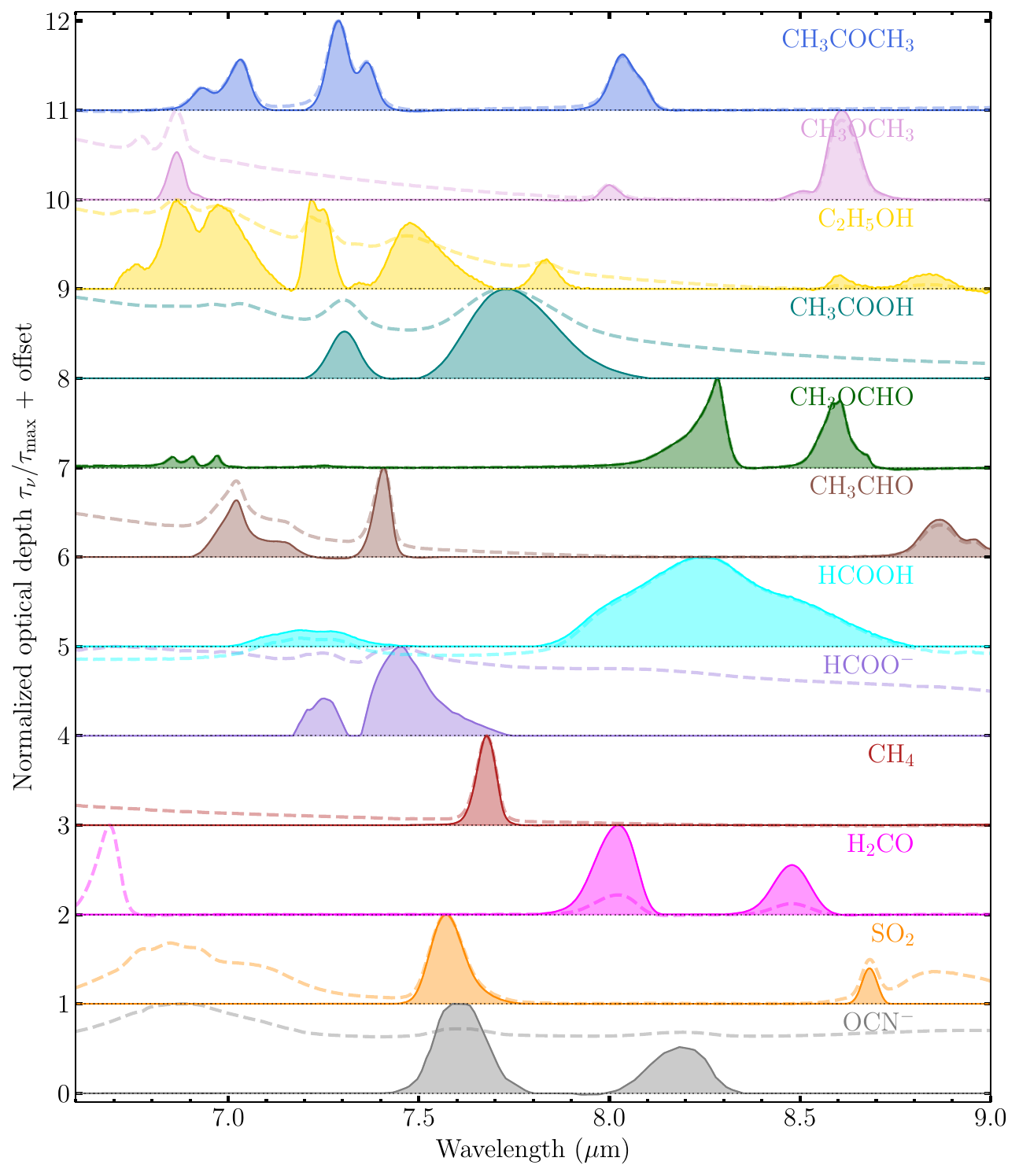}
\caption{Laboratory spectra before (dashed) and after (solid) baseline correction. Each spectrum is normalized to its peak optical depth with offsets added for all ice mixtures. Details on the laboratory spectra are summarized in Table \ref{tab:ice_references}.}
\label{fig:baseline_corr}
\end{figure} 

	Ice column densities are estimated from a comparison of the observed optical depth spectrum of IRAS\,18089 and laboratory spectra (Sect. \ref{sec:ice}). The properties of the ices and references of the ice spectra are summarized in Table \ref{tab:ice_references}. Figure \ref{fig:baseline_corr} shows laboratory spectra of pure ices or ice mixtures before and after baseline correction that were then used to estimate the ice column density towards the IRAS\,18089 mm peak position. The spectra shown in Fig. \ref{fig:baseline_corr} were normalized to their peak optical depth. The baseline correction includes removing contributions from other constituents in ice mixtures (e.g., H$_{2}$O and CH$_{3}$OH) as well as correcting offsets from zero optical depth in flat regions (e.g. HCOOH).

\section{Ice and gas-phase column densities}\label{app:coldens}
\begin{figure*}[!htb]
\centering
\includegraphics[width=1.0\textwidth]{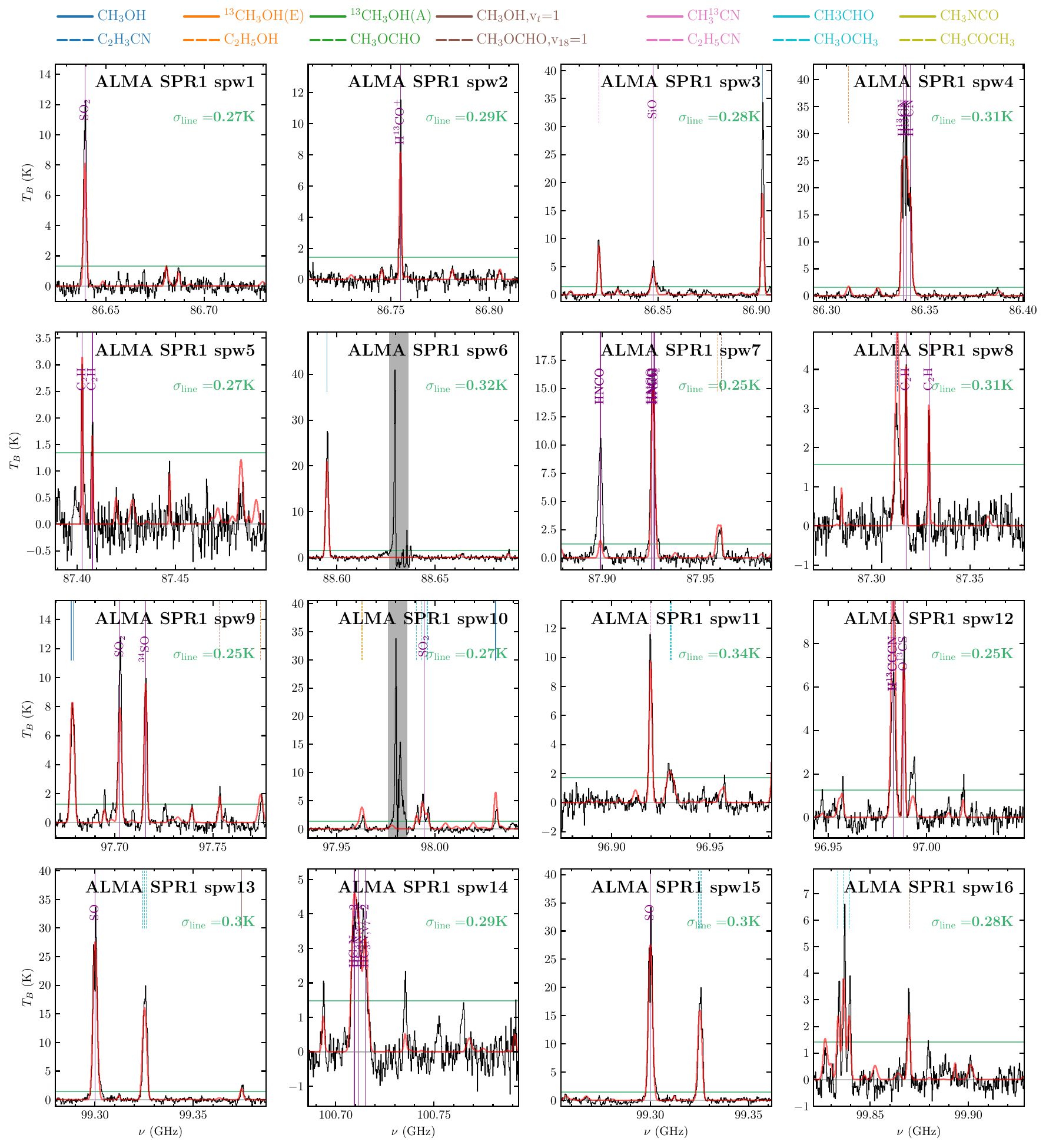}
\caption{ALMA 3\,mm spectra of SPR1 towards IRAS\,18089 mm. In all panels, the observed ALMA spectrum extracted from the mm peak position is shown in black. The best-fit model obtained with \texttt{xclass} is shown in red taking into account all molecules. The $5\sigma_\mathrm{line}$ level is highlighted by the horizontal green line. Vertical lines mark molecular transitions considered in the \texttt{xclass} fit with peak intensities $>5\sigma_\mathrm{line}$. Optically thick lines excluded from the fit are grey-shaded.}
\label{fig:alma}
\end{figure*} 

\begin{figure*}[!htb]
\centering
\includegraphics[width=1.0\textwidth]{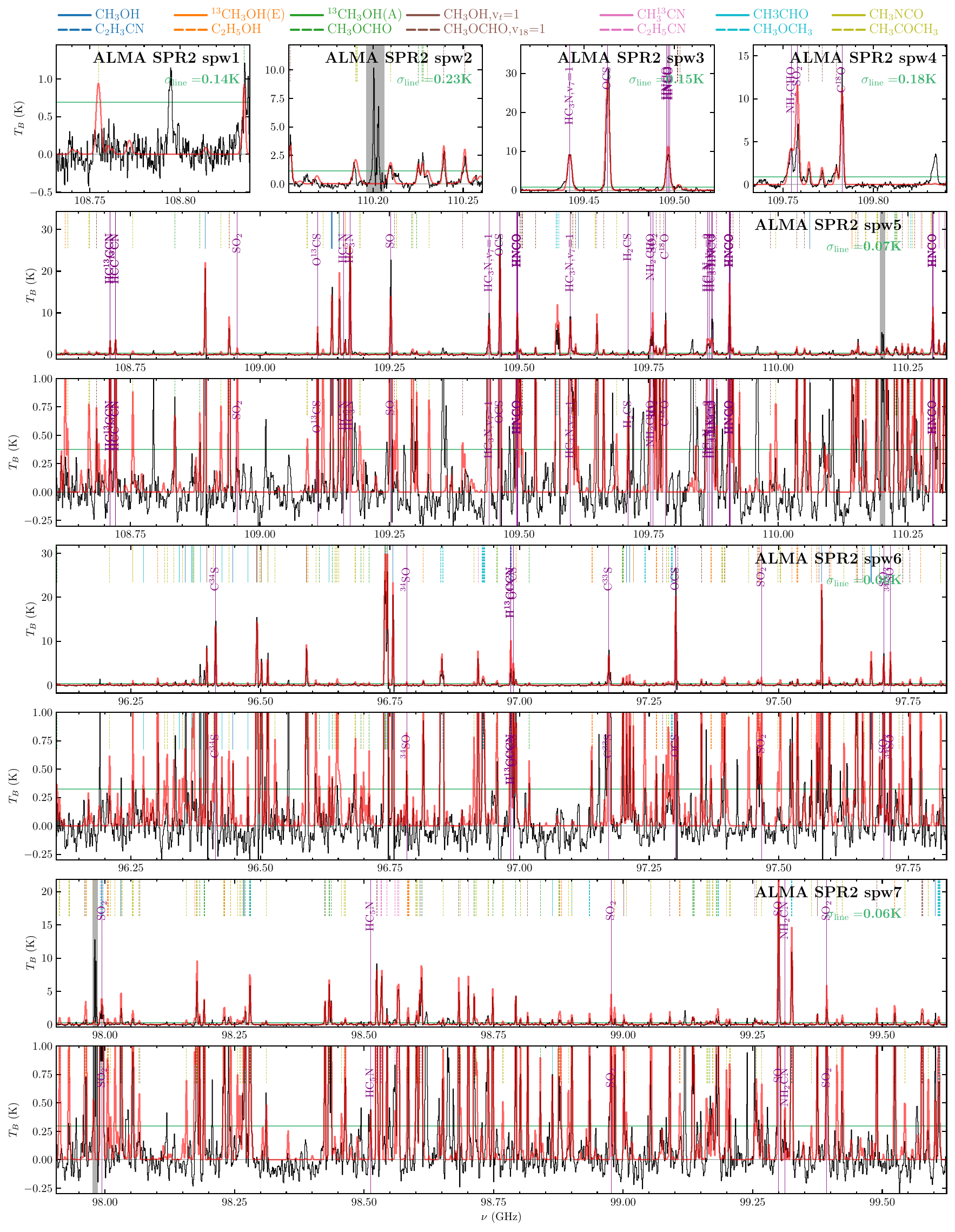}
\caption{The same as Fig. \ref{fig:alma}, but for SPR2. For the three broadband spws an additional zoom in panel is shown to highlight the fainter lines.}
\label{fig:alma2}
\end{figure*} 

\begin{figure*}[!htb]
\centering
\includegraphics[width=0.99\textwidth]{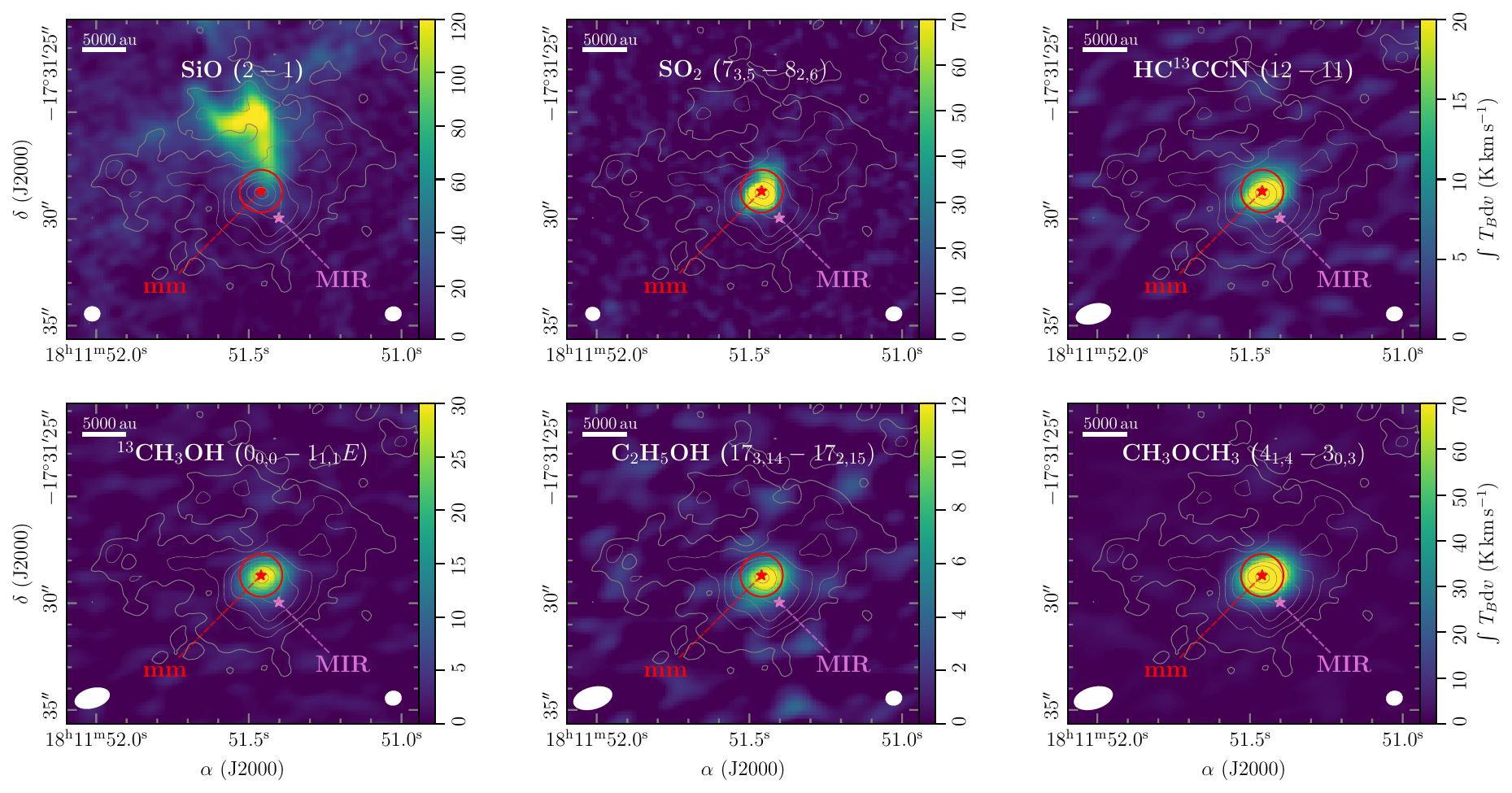}
\caption{Molecular line emission maps of IRAS\,18089. In all panels the ALMA line-integrated intensity ($\varv_\mathrm{LSR}\pm5$\,km\,s$^{-1}$) is shown in color and the grey contours are the ALMA 3\,mm continuum with steps from 5, 10, 20, 40, 80, 160, 320$\times\sigma_\mathrm{cont}$. The mm and MIR continuum peak positions are labeled and highlighted in red and pink. The red circle shows the aperture (1$''$ radius) used for spectra extraction towards the 3\,mm continuum peak. A scale bar in the top left panel marks a spatial scale of 5\,000\,au. The ellipse in the bottom left and bottom right corner highlights the angular resolution of the line and continuum data.}
\label{fig:ALMAline}
\end{figure*}

\begin{figure*}[!htb]
\centering
\includegraphics[width=0.99\textwidth]{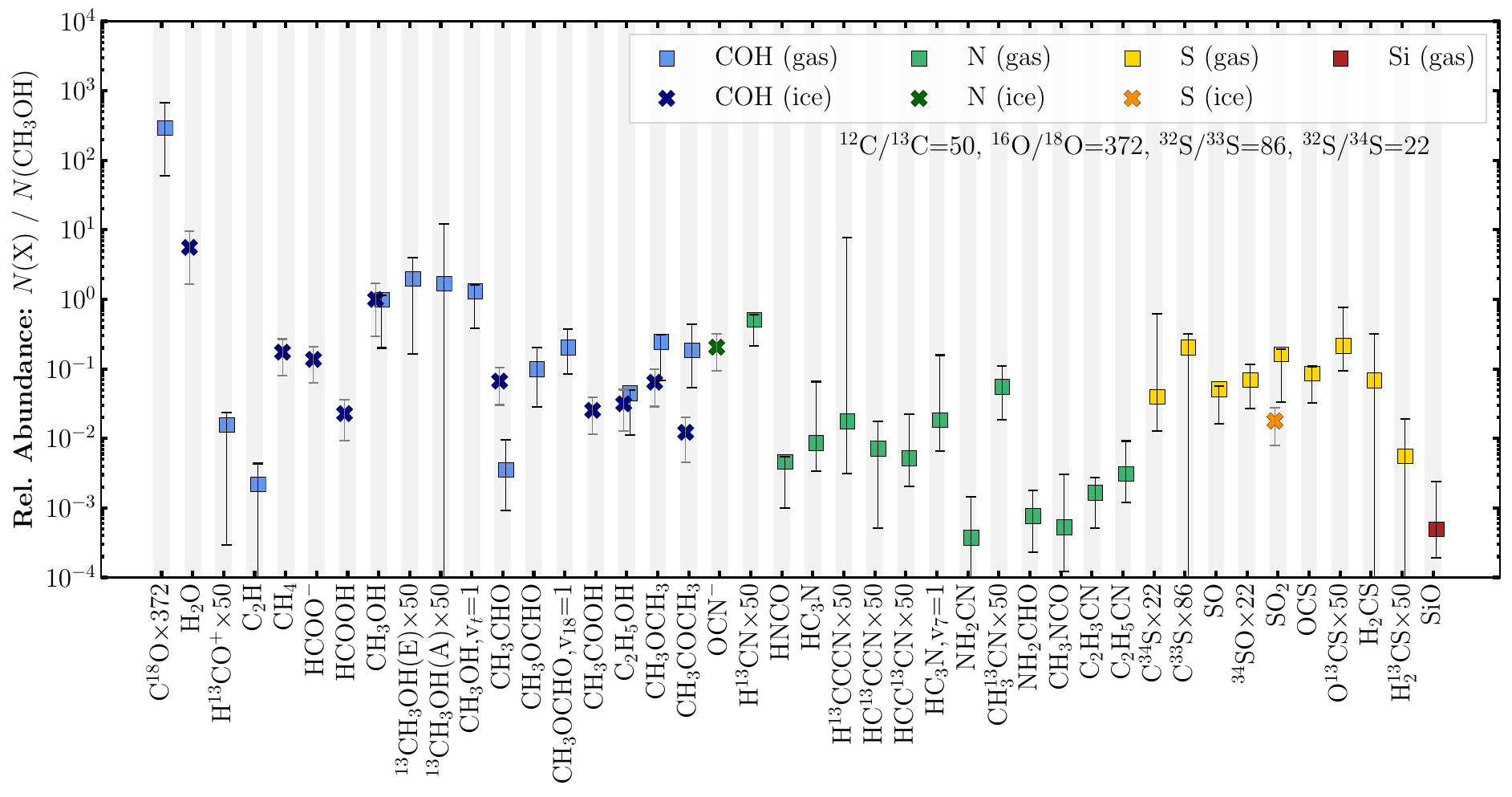}
\caption{The same as Fig. \ref{fig:comp_ice_gas}, but abundances are computed relative to CH$_{3}$OH.}
\label{fig:comp_ice_gas_CH3OH}
\end{figure*} 

	In Table \ref{tab:columdens} the column density of ice (Sect. \ref{sec:ice}) and gas-phase species (Sect. \ref{sec:gas}) are summarized derived using JWST and ALMA observations, respectively.

\setlength{\tabcolsep}{9pt}
\begin{table*}[!htb]
\caption{Molecular column densities in the ices and gas-phase towards IRAS\,18089.}
\label{tab:columdens}
\centering
\renewcommand{\arraystretch}{1.1}
\begin{tabular}{llcccccc}
\hline\hline
Molecule & \texttt{XCLASS} label (catalog) & \multicolumn{6}{c}{Column density}\\ 
 & & $N_\mathrm{gas}$ & $\Delta N_\mathrm{gas}^{\mathrm{low}}$ & $\Delta N_\mathrm{gas}^{\mathrm{upp}}$ & $N_\mathrm{ice}$ & $\Delta N_\mathrm{ice}^{\mathrm{low}}$ & $\Delta N_\mathrm{ice}^{\mathrm{upp}}$\\
 & & (cm$^{-2}$) & (cm$^{-2}$) & (cm$^{-2}$) & (cm$^{-2}$) & (cm$^{-2}$) & (cm$^{-2}$)\\
 \hline
C$^{18}$O$\times$372 & CO-18;v=0;(CDMS) & 5.9(20) & 3.3(20) & 7.5(20) & \ldots & \ldots & \ldots\\
H$_{2}$O & & \ldots & \ldots & \ldots & 1.5(19) & 7.3(18) & 7.3(18)\\
H$^{13}$CO$^{+}\times$50 & HC-13-O+;v=0;\#2(JPL) & 3.2(16) & 2.5(16) & 1.6(16) & \ldots & \ldots & \ldots\\
C$_{2}$H & CCH;v=0;(CDMS) & 4.4(15) & 3.5(15) & 4.3(15) & \ldots & \ldots & \ldots\\
CH$_4$ & & \ldots & \ldots & \ldots & 4.6(17) & 9.3(16) & 9.3(16)\\
HCOO$^{-}$ & & \ldots & \ldots & \ldots & 3.6(17) & 7.2(16) & 7.2(16)\\
HCOOH$^{(a)}$ & HCOOH;v=0;(CDMS) & $<$5.0(16) & \ldots & \ldots & 6.0(16) & 1.9(16) & 1.9(16)\\
CH$_{3}$OH$^{(b)}$ & CH3OH;v=0;\#2(JPL) & 2.0(18) & 1.1(18) & 1.9(17) & 2.6(18) & 1.3(18) & 1.3(18)\\
$^{13}$CH$_{3}$OH(E)$\times$50 & C-13-H3OH;v=0;E(CDMS) & 4.0(18) & 2.9(18) & 3.9(18) & \ldots & \ldots & \ldots\\
$^{13}$CH$_{3}$OH(A)$\times$50 & C-13-H3OH;v=0;A(CDMS) & 3.4(18) & 3.4(18) & 2.1(19) & \ldots & \ldots & \ldots\\
CH$_{3}$OH,v$_t$=1 & CH3OH;v12=1;\#2(JPL) & 2.7(18) & 1.1(18) & 5.5(17) & \ldots & \ldots & \ldots\\
CH$_{3}$CHO & CH3CHO;v=0;\#1(JPL) & 7.1(15) & 3.4(15) & 1.2(16) & 1.8(17) & 3.8(16) & 3.8(16)\\
CH$_{3}$OCHO & CH3OCHO;v=0;\#1(JPL) & 2.0(17) & 8.7(16) & 2.0(17) & \ldots & \ldots & \ldots\\
CH$_{3}$OCHO,v$_{18}$=1 & CH3OCHO;v18=1;(JPL) & 4.1(17) & 5.7(16) & 3.3(17) & \ldots & \ldots & \ldots\\
CH$_{3}$COOH$^{(a)}$ & CH3COOH;v=0;(CDMS) & $<$1.0(16) & \ldots & \ldots & 6.6(16) & 1.4(16) & 1.4(16)\\
C$_{2}$H$_5$OH & C2H5OH;v=0;\#1(CDMS) & 9.0(16) & 4.5(16) & 3.3(15) & 8.3(16) & 2.7(16) & 2.7(16)\\
CH$_{3}$OCH$_{3}$ & CH3OCH3;v=0;(CDMS) & 4.9(17) & 2.2(17) & 1.1(17) & 1.7(17) & 4.0(16) & 4.0(16)\\
CH$_{3}$COCH$_{3}$ & CH3COCH3;v=0;\#1(JPL) & 3.7(17) & 1.6(17) & 5.0(17) & 3.2(16) & 1.2(16) & 1.2(16)\\
OCN$^-$ & & \ldots & \ldots & \ldots & 5.4(17) & 1.1(17) & 1.1(17)\\
H$^{13}$CN$\times$50 & HC-13-N;v=0;hyp1(CDMS) & 1.0(18) & 1.7(17) & 1.3(17) & \ldots & \ldots & \ldots\\
HNCO & HNCO;v=0;(CDMS) & 9.2(15) & 5.0(15) & 1.4(15) & \ldots & \ldots & \ldots\\
HC$_{3}$N & HCCCN;v=0;\#2(JPL) & 1.7(16) & 3.9(15) & 1.1(17) & \ldots & \ldots & \ldots\\
H$^{13}$CCCN$\times$50 & HC-13-CCN;v=0;\#2(JPL) & 3.6(16) & 2.1(16) & 1.6(19) & \ldots & \ldots & \ldots\\
HC$^{13}$CCN$\times$50 & HCC-13-CN;v=0;\#2(JPL) & 1.4(16) & 1.1(16) & 2.0(16) & \ldots & \ldots & \ldots\\
HCC$^{13}$CN$\times$50 & HCCC-13-N;v=0;\#2(JPL) & 1.0(16) & 2.4(15) & 3.4(16) & \ldots & \ldots & \ldots\\
HC$_{3}$N,v$_7$=1 & HCCCN;v7=1;(CDMS) & 3.7(16) & 1.2(16) & 2.8(17) & \ldots & \ldots & \ldots\\
NH$_{2}$CN & NH2CN;v=0;(JPL) & 7.5(14) & 6.0(14) & 2.1(15) & \ldots & \ldots & \ldots\\
CH$_{3}^{13}$CN$\times$50 & CH3C-13-N;v=0;\#2(JPL) & 1.1(17) & 4.0(16) & 1.1(17) & \ldots & \ldots & \ldots\\
NH$_{2}$CHO & HC(O)NH2;v=0;(CDMS) & 1.6(15) & 6.6(14) & 2.0(15) & \ldots & \ldots & \ldots\\
CH$_{3}$NCO & CH3NCO;v=0;\#1(CDMS) & 1.1(15) & 5.6(14) & 5.0(15) & \ldots & \ldots & \ldots\\
C$_{2}$H$_{3}$CN & C2H3CN;v=0;\#2(JPL) & 3.3(15) & 1.3(15) & 2.1(15) & \ldots & \ldots & \ldots\\
C$_{2}$H$_{5}$CN & C2H5CN;v=0;(CDMS) & 6.2(15) & 1.5(15) & 1.2(16) & \ldots & \ldots & \ldots\\
C$^{34}$S$\times$22 & CS-34;v=0;(CDMS) & 8.0(16) & 3.1(16) & 1.2(18) & \ldots & \ldots & \ldots\\
C$^{33}$S$\times$86 & CS-33;v=0;(CDMS) & 4.1(17) & 3.4(17) & 2.2(17) & \ldots & \ldots & \ldots\\
SO & SO;v=0;\#2(CDMS) & 1.0(17) & 3.9(16) & 2.6(15) & \ldots & \ldots & \ldots\\
$^{34}$SO$\times$22 & S-34-O;v=0;(CDMS) & 1.4(17) & 3.3(16) & 9.1(16) & \ldots & \ldots & \ldots\\
SO$_{2}$ & SO2;v=0;(CDMS) & 3.3(17) & 1.8(17) & 5.1(16) & 4.7(16) & 1.2(16) & 1.2(16)\\
OCS & OCS;v=0;(CDMS) & 1.7(17) & 4.7(16) & 4.1(16) & \ldots & \ldots & \ldots\\
O$^{13}$CS$\times$50 & OC-13-S;v=0;(CDMS) & 4.3(17) & 2.2(16) & 1.1(18) & \ldots & \ldots & \ldots\\
H$_{2}$CS & H2CS;v=0;\#2(JPL) & 1.4(17) & 1.4(17) & 5.0(17) & \ldots & \ldots & \ldots\\
H$_{2}^{13}$CS$\times$50 & H2C-13-S;v=0;\#2(JPL) & 1.1(16) & 1.0(16) & 2.7(16) & \ldots & \ldots & \ldots\\
SiO & SiO;v=0;\#1(CDMS) & 9.9(14) & 2.4(14) & 3.8(15) & \ldots & \ldots & \ldots\\
\hline
\end{tabular}
\tablefoot{The format of the column density column with its upper and lower uncertainty ($N^{+\Delta N^{\mathrm{upp}}}_{-\Delta N^{\mathrm{low}}}$) is $a$($b$)=$a\times10^{b}$. The column density of less abundant isotopologues were converted to their corresponding main isotopologue (Sect. \ref{sec:gas}). For each species the \texttt{XCLASS} molecule label is listed as well as the reference catalog, either the Cologne Database for Molecular Spectroscopy \citep[CDMS,][]{CDMS} or the Jet Propulsion Laboratory \citep[JPL,][]{JPL}. The \texttt{XCLASS} database version as of December 2024 was used in this work.\\ 
$^{(a)}$: Upper limits estimated using $T_\mathrm{rot}$=100\,K and $\Delta \varv$=5\,km\,s$^{-1}$ (mean values derived from detected species).\\
$^{(b)}$: The CH$_{3}$OH ice column density was inferred indirectly from the MIR source measurements \citep{vanDishoeck2025} assuming that the CH$_{3}$OH/H$_{2}$O column density ratio is equal for the mm and MIR sources (Sect. \ref{sec:ice}).
}
\end{table*}

\section{Additional tests of the ice fitting approach}\label{app:testsice}

	In Sect. \ref{sec:ice} we derive column densities of molecular ices detected in the JWST/MIRI-MRS spectrum towards the IRAS\,18089 hot core. We model the SED of the continuum, including dust and water ice absorption (Fig. \ref{fig:spectrumIR}). The spectrum is then further corrected by removing gas-phase SO$_{2}$ emission and CH$_{4}$ absorption lines between 7.2\,$\upmu$m and 7.7\,$\upmu$m and a local continuum is estimated between 6.6\,$\upmu$m and 9\,$\upmu$m by using a 3rd order polynomial spline interpolation (Fig. \ref{fig:spectrumtau}). The ice contributions and column densities are then evaluated based on a least squares fit (Fig. \ref{fig:spectrumICE}). We refer to this ice fitting setup as the ``Final fit'' setup in the following.
	
	Given the amount of steps required before the ice fitting itself, we provide here additional tests to demonstrate that our derived ice column densities presented in Sect. \ref{sec:ice} are robust. All changes mentioned in the following test setups, are relative to the ``Final fit'' setup. For Test A, we only fit the optical depth spectrum up to 8.8\,$\upmu$m, instead of 9\,$\upmu$m, since the noise is significantly increasing after 8.8\,$\upmu$m and then also reaching the noise floor at 9\,$\upmu$m (Fig. \ref{fig:spectrumIR}). For Test B we use the results from the least squares fit and use them as a prior for an MCMC fit using the \texttt{emcee} package \citep{ForemanMackey2013}. For Test C we use a 5th order instead of a 3rd order polynomial spline to estimate the local continuum (Fig. \ref{fig:spectrumtau}). For Test D we perform our ice analysis based on the IRAS\,18089 spectrum without a correction of the SO$_{2}$ and CH$_{4}$ gas-phase lines.
	
\begin{figure*}[!htb]
\centering
\includegraphics[width=0.45\textwidth]{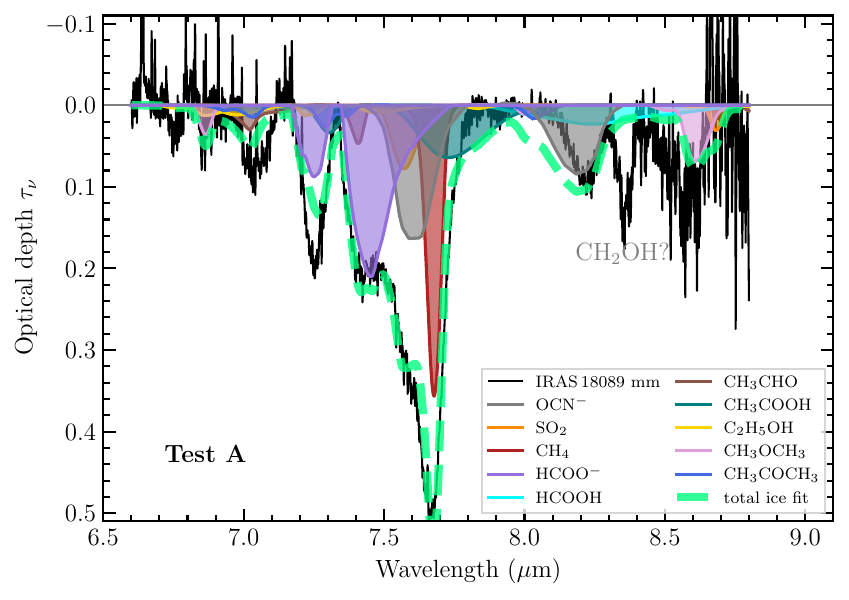}
\includegraphics[width=0.45\textwidth]{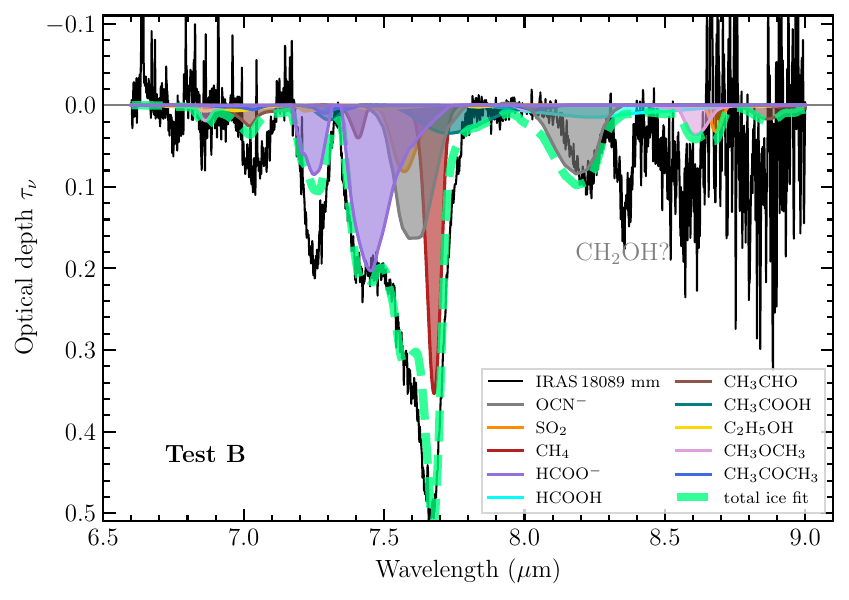}
\includegraphics[width=0.45\textwidth]{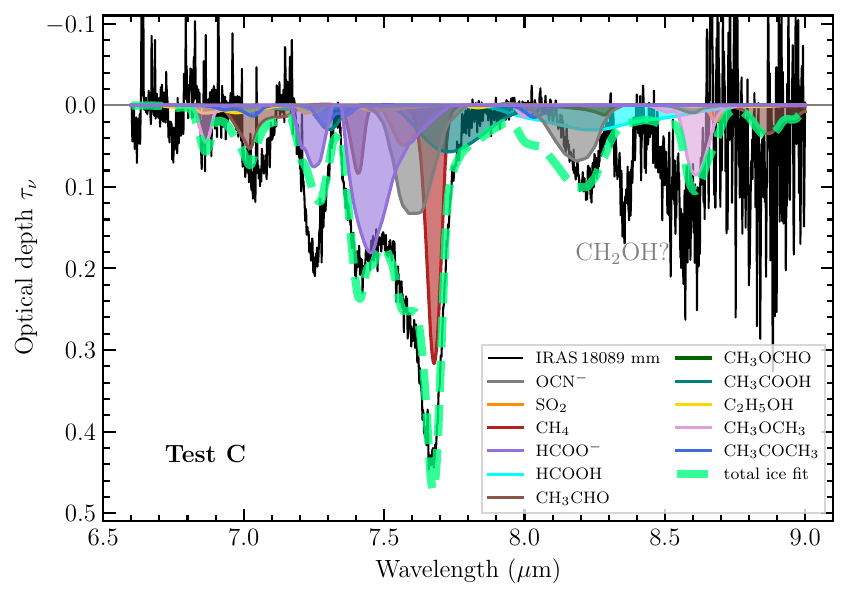}
\includegraphics[width=0.45\textwidth]{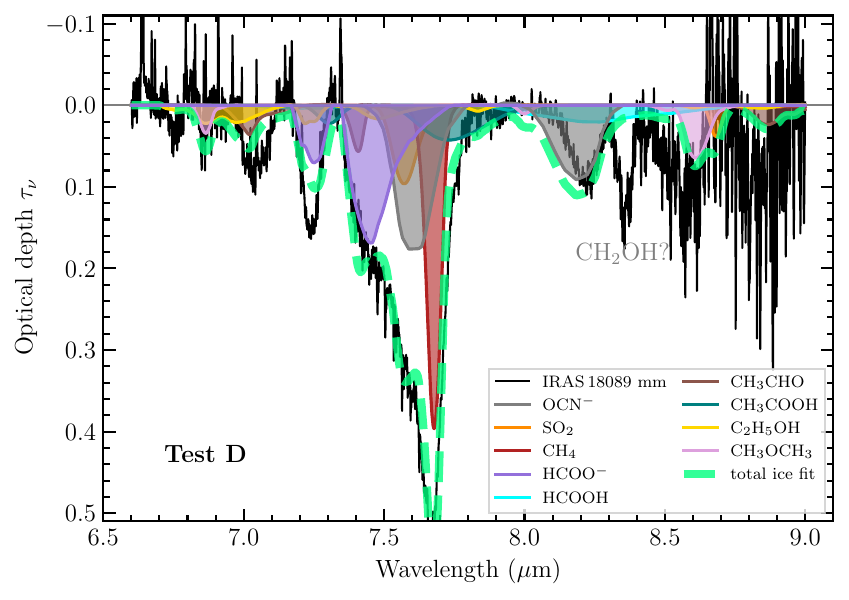}
\caption{The same as Fig. \ref{fig:spectrumICE}, but including different tests for the ice fitting method.}
\label{fig:spectrumICE_test}
\end{figure*}

\begin{figure*}[!htb]
\centering
\includegraphics[width=0.67\textwidth]{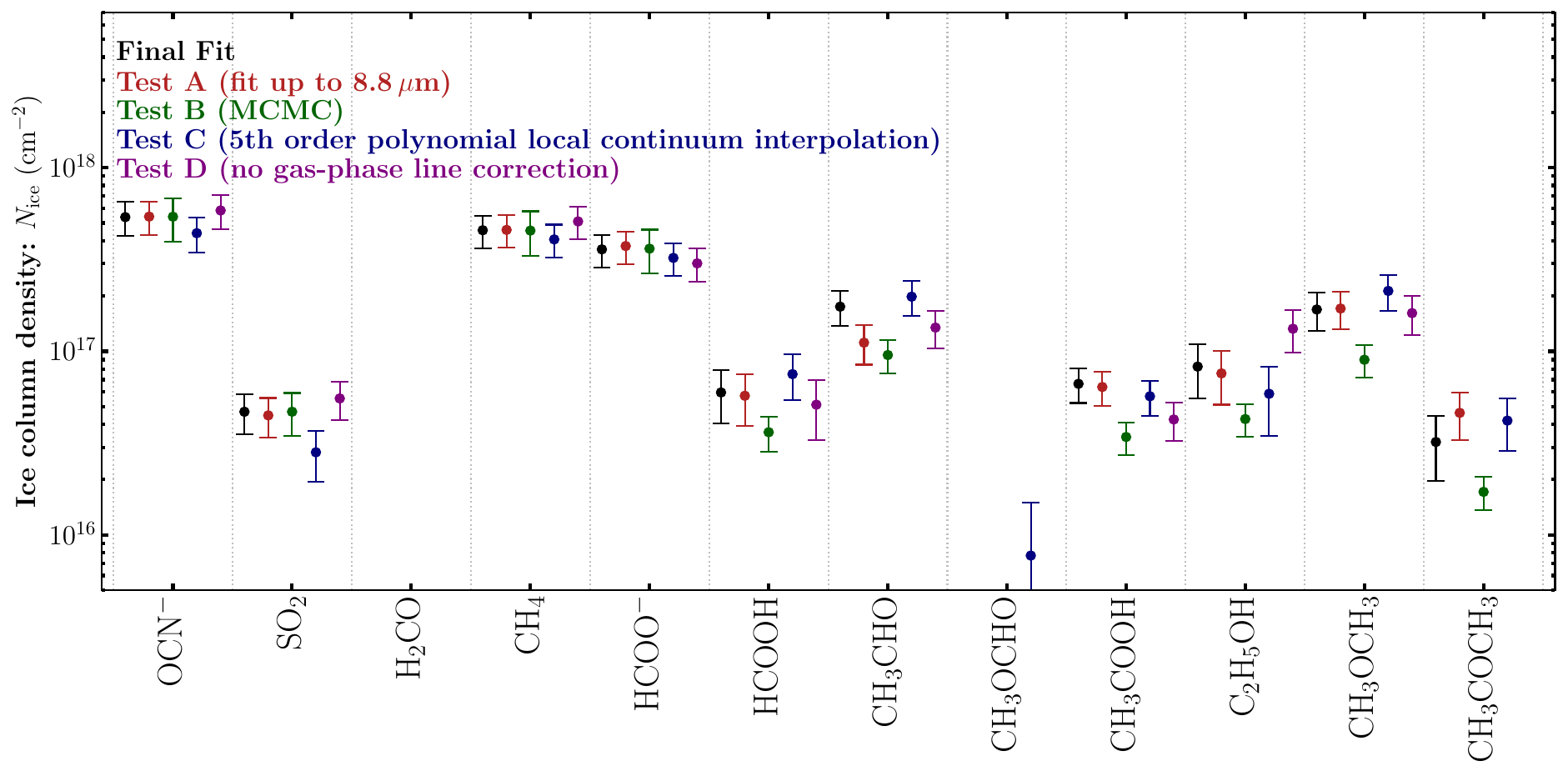}
\caption{Comparison of the derived ice column densities of the final fit (Fig. \ref{fig:spectrumICE}) and additional tests (Fig. \ref{fig:spectrumICE_test}).}
\label{fig:test_comparison}
\end{figure*}

	The best-fit ice spectra and column density results of these tests are presented in Fig. \ref{fig:spectrumICE_test} and Table \ref{tab:columdens_test}. A comparison of the ice column density of the final fit setup and these tests is shown in Fig. \ref{fig:test_comparison}. In order to compare the different fit results, we compute the Residual sum of squares (RSS) value for all tests which is presented in Table \ref{tab:columdens_test}. For the final fit setup, we obtain RSS=4.7.
	
	Reducing the fit range from 9\,$\upmu$m to 8.8\,$\upmu$m (Test A) only affects the CH$_{3}$CHO ice column density for which the column density is then slightly lower. However given that CH$_{3}$CHO has ice bands present between 8.8\,$\upmu$m and 9\,$\upmu$m (Fig. \ref{fig:baseline_corr}), including this range is important to best constrain this ice species.
	
\begin{figure*}[!htb]
\centering
\includegraphics[width=0.9\textwidth]{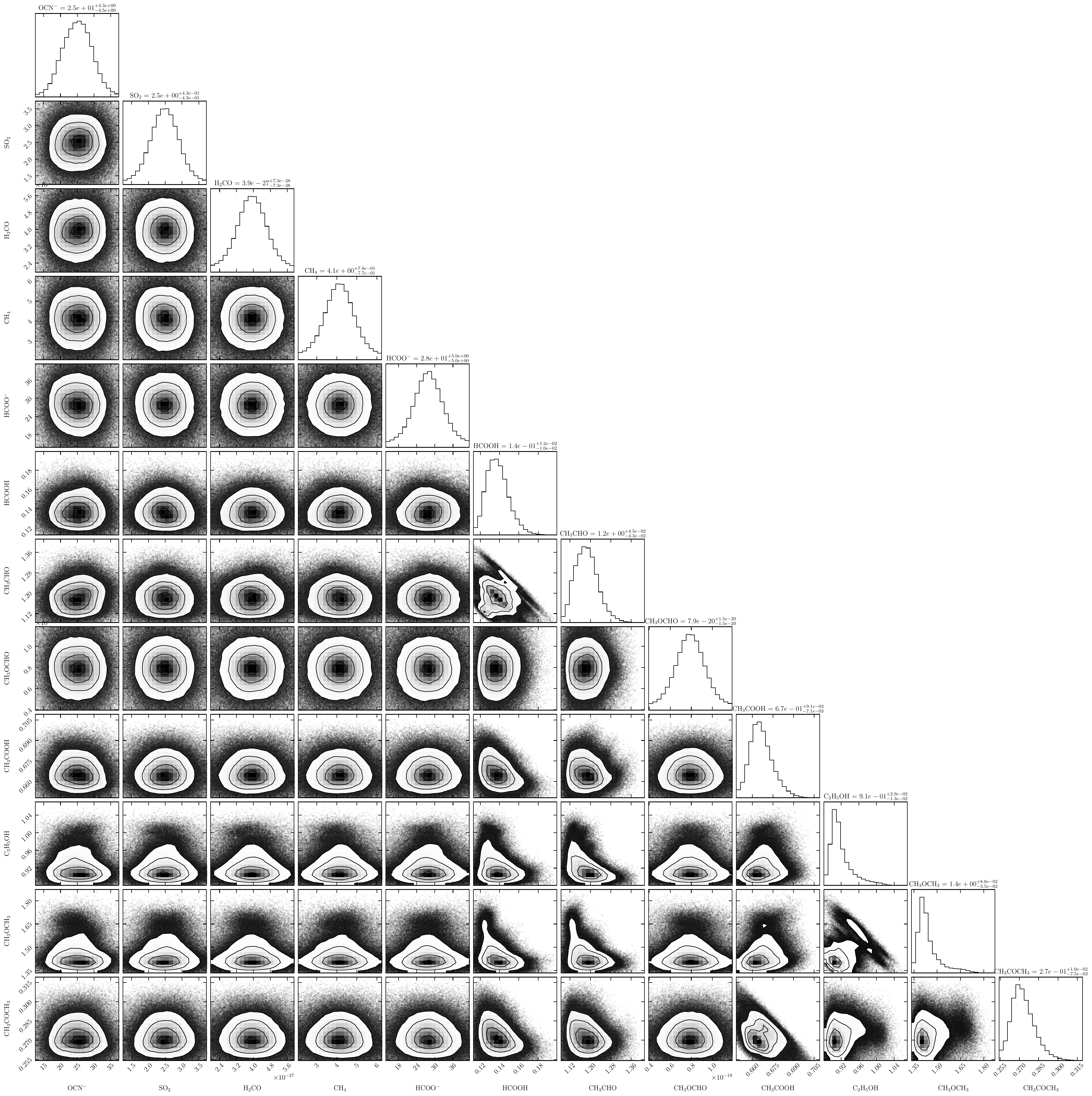}
\caption{MCMC fit of the IRAS\,18089 ice spectrum (Fig. \ref{fig:spectrumICE}) using the \texttt{emcee} package. The best-fit scaling factor (Eq. \ref{eq:icefit}) is shown for each ice species. The corresponding ice column densities are summarized in Table \ref{tab:columdens_test}.}
\label{fig:MCMC_ice}
\end{figure*}

	To further explore the parameter space of the 12 considered ice species with MCMC (Test B), we use 500 walkers, 35\,000 iterations (7\,000 as burn-in) and as a prior the best-fit from the least squares optimization. The MCMC best-fit has lower column densities for many species, including HCOOH, CH$_{3}$CHO, CH$_{3}$COOH, C$_{2}$H$_{5}$OH, CH$_{3}$OCH$_{3}$, and CH$_{3}$COCH$_{3}$. Despite using the least squares optimization as a prior, the MCMC walkers get stuck in local minima resulting in a worse RSS (5.2) compared to the least squares fit (4.7) which can be attributed to the large number of free parameters (12). Hence, the non-linear least squares fit with \texttt{curve\_fit} provides better fit results. The corner plot of the MCMC ice fitting is presented in Fig. \ref{fig:MCMC_ice}. We find small degeneracies between HCOOH and CH$_{3}$CHO as well as C$_{2}$H$_{5}$OH and CH$_{3}$OCH$_{3}$ ice column densities. In contrast to that, \citet{Chen2024} found a degeneracy between the HCOOH and CH$_{3}$OCHO ice column densities (their Fig. K.2). This suggests that potential degeneracies of different ice species depend heavily on the analyzed source.
		
	The local continuum interpolation is not straightforward (Test C) and does influence heavily the estimated ice column densities, especially SO$_2$ \citep{Gross2026}. Within the uncertainties we find significant deviations from our final fit results for SO$_{2}$ and CH$_{3}$OCHO. For SO$_{2}$ we find differences of about a factor of two. While CH$_{3}$OCHO has no significant contributions in the final fit, we find a minor contribution in the Test C model. In agreement with \citet{Gross2026} we find that the SO$_{2}$ ice column density is influenced by the choice of the local continuum and in addition we find that CH$_{3}$OCHO also is affected. For the remaining ice species we find consistent results using both local continuum interpolations.

	The removal of gas-phase SO$_{2}$ and CH$_{4}$ lines (Test D) is essential when estimating ice column densities, as strong transitions overlap with 7.2\,$\upmu$m and 7.4\,$\upmu$m ice absorption bands (Fig. \ref{fig:spectrumICE_test}). Within the uncertainties, we find significant differences for CH$_3$COOH, C$_2$H$_5$OH, and CH$_3$COCH$_3$ ice column densities.

\setlength{\tabcolsep}{3pt}
\begin{table*}[!htb]
\caption{Molecular column densities in the ices towards IRAS\,18089 using different test setups.}
\label{tab:columdens_test}
\centering
\renewcommand{\arraystretch}{1.5}
\begin{tabular}{l|ccc|ccc|ccc|ccc}
\hline\hline
 & & Test A & & & Test B & & & Test C & & & Test D & \\
 & \multicolumn{3}{c|}{fit range} & \multicolumn{3}{c|}{MCMC} & \multicolumn{3}{c|}{local continuum} & \multicolumn{3}{c}{no gas-phase correction}\\
Molecule & $N_\mathrm{ice}$ & $\Delta N_\mathrm{ice}^{\mathrm{low}}$ & $\Delta N_\mathrm{ice}^{\mathrm{upp}}$ & $N_\mathrm{ice}$ & $\Delta N_\mathrm{ice}^{\mathrm{low}}$ & $\Delta N_\mathrm{ice}^{\mathrm{upp}}$ & $N_\mathrm{ice}$ & $\Delta N_\mathrm{ice}^{\mathrm{low}}$ & $\Delta N_\mathrm{ice}^{\mathrm{upp}}$ & $N_\mathrm{ice}$ & $\Delta N_\mathrm{ice}^{\mathrm{low}}$ & $\Delta N_\mathrm{ice}^{\mathrm{upp}}$\\
 & (cm$^{-2}$) & (cm$^{-2}$) & (cm$^{-2}$) & (cm$^{-2}$) & (cm$^{-2}$) & (cm$^{-2}$) & (cm$^{-2}$) & (cm$^{-2}$) & (cm$^{-2}$) & (cm$^{-2}$) & (cm$^{-2}$) & (cm$^{-2}$)\\
 \hline
OCN$^{-}$ & 5.4(17) & 1.1(17) & 1.1(17) & 5.4(17) & 1.4(17) & 1.4(17) & 4.4(17) & 9.4(16) & 9.4(16) & 5.8(17) & 1.2(17) & 1.2(17) \\
SO$_2$ & 4.5(16) & 1.1(16) & 1.1(16) & 4.7(16) & 1.2(16) & 1.2(16) & 2.8(16) & 8.8(15) & 8.8(15) & 5.5(16) & 1.3(16) & 1.3(16) \\
H$_2$CO & \ldots & \ldots & \ldots & \ldots & \ldots & \ldots & \ldots & \ldots & \ldots & \ldots & \ldots & \ldots \\
CH$_4$ & 4.6(17) & 9.3(16) & 9.3(16) & 4.5(17) & 1.2(17) & 1.2(17) & 4.1(17) & 8.3(16) & 8.3(16) & 5.1(17) & 1.0(17) & 1.0(17) \\
HCOO$^{-}$ & 3.7(17) & 7.5(16) & 7.5(16) & 3.6(17) & 9.7(16) & 9.8(16) & 3.2(17) & 6.5(16) & 6.5(16) & 3.0(17) & 6.1(16) & 6.1(16) \\
HCOOH & 5.7(16) & 1.8(16) & 1.8(16) & 3.6(16) & 7.7(15) & 7.9(15) & 7.5(16) & 2.1(16) & 2.1(16) & 5.1(16) & 1.8(16) & 1.8(16) \\
CH$_3$CHO & 1.1(17) & 2.7(16) & 2.7(16) & 9.5(16) & 1.9(16) & 1.9(16) & 2.0(17) & 4.2(16) & 4.2(16) & 1.3(17) & 3.1(16) & 3.1(16) \\
CH$_3$OCHO & \ldots & \ldots & \ldots & \ldots & \ldots & \ldots & 7.7(15) & 7.2(15) & 7.2(15) & \ldots & \ldots & \ldots \\
CH$_3$COOH & 6.4(16) & 1.4(16) & 1.4(16) & 3.4(16) & 6.8(15) & 6.8(15) & 5.7(16) & 1.2(16) & 1.2(16) & 4.2(16) & 9.9(15) & 9.9(15) \\
C$_2$H$_5$OH & 7.6(16) & 2.5(16) & 2.5(16) & 4.3(16) & 8.6(15) & 8.7(15) & 5.9(16) & 2.4(16) & 2.4(16) & 1.3(17) & 3.4(16) & 3.4(16) \\
CH$_3$OCH$_3$ & 1.7(17) & 3.9(16) & 3.9(16) & 9.0(16) & 1.8(16) & 1.9(16) & 2.1(17) & 4.8(16) & 4.8(16) & 1.6(17) & 3.9(16) & 3.9(16) \\
CH$_3$COCH$_3$ & 4.6(16) & 1.3(16) & 1.3(16) & 1.7(16) & 3.5(15) & 3.5(15) & 4.2(16) & 1.3(16) & 1.3(16) & \ldots & \ldots & \ldots\\
\hline
Total fit RSS & & 4.7 & & & 5.2 & & & 4.6 & & & 4.9 & \\
\hline
\end{tabular}
\tablefoot{The format of the column density with its upper and lower uncertainty ($N^{+\Delta N^{\mathrm{upp}}}_{-\Delta N^{\mathrm{low}}}$) is $a$($b$)=$a\times10^{b}$. The ice column densities of the final fit are summarized in Table \ref{tab:columdens} resulting in RSS of 4.7.}
\end{table*}

\end{appendix}

\end{document}